\documentclass[journal,draftcls,onecolumn,english]{IEEEtran}
\usepackage{amsmath,amssymb,amsfonts}
\usepackage{graphicx}
\usepackage{textcomp}
\usepackage{multirow}
\usepackage{blindtext}
\usepackage{algorithm,algpseudocode}

\def\BibTeX{{\rm B\kern-.05em{\sc i\kern-.025em b}\kern-.08em
    T\kern-.1667em\lower.7ex\hbox{E}\kern-.125emX}}

\begin{document}

\title{QSMVM: QoS-aware and social-aware multimetric routing protocol for video-streaming services over MANETs}

\author{\IEEEauthorblockN{Efra\'in Palacios Jara}.
\IEEEauthorblockA{\textit{Department of Networking Engineering. Universitat Politècnica de Catalunya}
\and
\IEEEauthorblockN{Ahmad Mohamad Mezhe}.
\IEEEauthorblockA{Department of Networking Engineering. Universitat Politècnica de Catalunya}
\and
\IEEEauthorblockN{M\'onica Aguilar Igartua}.
\IEEEauthorblockA{Department of Networking Engineering. Universitat Politècnica de Catalunya. monica.aguilar@upc.edu}\\
\and
\IEEEauthorblockN{Rebeca P. D\'iaz Redondo}.
\IEEEauthorblockA{atlanTTic Research Centre. University of Vigo. rebeca@det.uvigo.es  \\}
\and
\IEEEauthorblockN{Ana Fern\'andez Vilas}.
\IEEEauthorblockA{atlanTTic Research Centre. University of Vigo. avilas@det.uvigo.es\\}
}}

\maketitle

\abstract{A mobile ad hoc network (MANET) is a set of autonomous mobile devices connected by wireless links in a distributed manner and without a fixed infrastructure. Real-time multimedia services, such as video-streaming over MANETs, offers very promising applications, e.g. two members of a group of tourists who want to share a video transmitted through the MANET they form; a video-streaming service deployed over a MANET where users watch a film; among other examples. On the other hand, social web technologies, where people actively interact online with others through social networks, are leading to a socialization of networks. Information of interaction among users is being used to provide socially-enhanced software. To achieve this, we need to know the strength of the relationship between a given user and each user they interact with. This strength of the relationship can be measured through a concept called tie strength (TS), first introduced by Mark Granovetter in 1973. In this article, we modify our previous proposal named multipath multimedia dynamic source routing (MMDSR) protocol to include a social metric TS in the decisions taken by the forwarding algorithm. We find a trade-off between the quality of service (QoS) and the trust level between users who form the forwarding path in the MANET. Our goal is to increase the trust metric while the QoS is not affected significantly.}


\section{Introduction}

A mobile ad hoc network (MANET) is a set of autonomous mobile devices connected by wireless links that cooperate with each other to achieve network functionalities in a distributed way and without a fixed infrastructure. Since mobile devices are free to move arbitrarily, this makes the network topology change dynamically \cite{Falko_book} \cite{Azzedine_book}.

This work is based on our previous proposal named \textit{multipath multimedia dynamic source routing} (MMDSR), which was previously presented in \cite{Ahmad_Jitel_2015} and \cite{GT_video_MANET_Ahmad}. The goal of this article is to further improve our proposal considering the social dimension by including a new \textit{tie strength} (TS) metric as a  measure of the interaction among users in social networks. As a result, our new proposal named \textit{QoS-aware and social-aware multimetric routing protocol for video-streaming services over MANETs} (QSMVM) is able to offer a good trade-off between QoS and confidence in the forwarding nodes involved in the path.

Providing multimedia services over MANETs represent a very attractive application of wireless networks for many cases, for instance in a disaster scenario or for entertainment purposes. The growth and big interest in ad hoc wireless networks has brought the development of new routing protocols and applications in many research areas such as governmental or commercial sector, where ad hoc networks can be used in emergency or rescue operations for natural disasters (an earthquake, flood, etc.) where it is very likely the network infrastructure get destroyed \cite{QoS_disaster}. Also, video-streaming services allow citizens to report any incident in the city (e.g., a traffic accident or a traffic jam), so that city authorities could have a better idea of the situation with a short and light video than just with a text message \cite{INRISCO}. Another example for entertainment purposes is a video-streaming service to watch movies over MANETs \cite{QoS_video1}, \cite{QoS_video2}.

On the other hand, social web technologies, where people actively interact online with others through social networks, share multimedia contents, post comments and so on, are leading to a socialization of networks. Nowadays, users not only ask for information but also produce it. Currently, information of interaction among users in social networks is used to provide socially-enhanced software, always aware of the social environment of the users. To achieve this, we need to know the strength of the relationship between a given user and each user they interact with. The concept of tie strength (TS) was firstly introduced by Granovetter in 1973 \cite{Granovetter73} as a function of duration, emotional intensity, intimacy and exchange of services. He distinguished two kinds of ties in social networks: strong and weak ties \cite{Servia13}. The TS concept summarizes the idea of social closeness between human beings. We claim that TS could also be used as a measure of trust in network communications.  This aspect is relevant when information travels throughout different actors/elements, since trust (or closeness) may be used to assure the information travels through actors or elements that the two ends of the communication (i.e., sender and receiver) trust. Thus, considering together a social metric like the TS along with traditional QoS metrics (e.g., packet losses, packet delay, bandwidth...) in the design of routing protocols, can offer a clear benefit to individuals concerned not only by the QoS level but also by the privacy of their communications.

\textbf{Motivation of our proposal:}
Our main motivation is to jointly consider several QoS metrics together with a new social metric in the design of a multimetric routing protocol for MANETs. The goal is to provide video-streaming services over MANETs improving not only the QoS offered to the end user, but also improving the trust on the intermediate forwarding nodes along the path from source to destination. Thus, our multimetric routing protocol for MANETs focuses on maximizing both the QoS and the trust of the forwarding path.

Our main contributions can be summarized as follows:
\begin{itemize}
\item We aim to exploit the aforementioned information of interaction among users and develop a QoS-aware and TS-aware routing protocol to provide video-streaming services over MANETs. Our routing protocol considers both the relationship among users while providing a good performance in terms of quality of service.
\item We claim that this idea can be useful to deal with possible network security trust issues in MANETs, since with our approach users’ information is routed through paths formed by users with high interaction and trust with each other. Therefore, with our proposal, packets are preferably forwarded by trusted nodes rather than by strangers that are not trusted, which leads to increased trust in communication.
\item In this work  we find a trade-off between (i) the user performance in terms of QoS parameters (focusing mainly in the percentage of packet losses and the end-to-end packet delay) and (ii) a good level of trust or confidence from the user’s point of view, represented by the average TS of the forwarding nodes.
\item We include simulation results under different scenarios that show the benefits of our proposal.
\end{itemize}

The rest of the paper is structured as follows. Section \ref{sec:basics} introduce the basics of our framework. In Section \ref{sec:QoS_routing} we summarize the main features of our proposal of a QoS-aware multimetric routing protocol. Section \ref{sec:TS} gives a brief explanation of the TS concept and its application in a social-aware routing protocol to select the best forwarding path. Simulation results are shown and analyzed in section \ref{sec:simulations}. Finally, conclusions and future work are given in section \ref{sec:conclusions}. 

\section{Related work}
\label{sec:related_work}
In this section we highlight some related works from recent years abut the topics related to our proposal: (i) video-streaming services over MANETs, (ii) QoS-aware multimetric routing for MANETs, (iii) social MANETs.

(i) There are several proposals about \textit{video-streaming services over MANETs} in the last decade. The work \cite{data_rate_video_manet} introduces an optimal rate control for video stream transmission to minimize both the client buffer utilization and the video transmission rate. The optimal sending data rate from server is analytically derived and the buffer occupancy is numerically calculated.
The approach presented in \cite{p2p_video_manet} chooses the better quality links for routing instead of the minimum hop-count path. Then, they distribute the video-streaming to receivers by using multicasting in multi-channel Wireless Mesh Networks. They design a multicast version of the well-known AODV routing protocol to construct two disjoint multicast trees as the backbone for a peer-to-peer structure. Finally, they adopt the multiple description coding scheme to encode the video into two independent sub-streams and transmit them separately along those multicast trees. NS-2 simulations show that in higher traffic load environment, their scheme reduces latency and improves the packet delivery ratio. In \cite{QoS_Greco}, the authors propose a cross-layer congestion control strategy where the MAC layer is video-coding aware and adjusts its transmission parameters via congestion/distortion optimization, showing a gain in terms of PSNR and delay reduction. The authors in \cite{QoS_video3} propose a multi-objective function that minimizes the number of packets injected in the network and maximizes the path diversity among the different video encoding descriptions. They also include a cross-layer congestion control strategy where the MAC layer is video-coding aware, achieving a consistent gain in terms of both PSNR and delay reduction. The authors in \cite{GT_video_MANET_Ahmad} propose a multipath game-theoretical routing protocol to transmit video-warning messages over MANETs. The considered scenario is a video-warning service in smart cities, so that when an accident happens, dynamic sensors (e.g., citizens with smart phones or tablets, smart vehicles and buses) shoot a video clip of the accident and send it through the MANET. Instead of sending video frames always through the best available path, users \textit{play} a strategic routing game according which video frames are sent through one of the two best paths according to a probability $p$ that varies on some network features (e.g., number of nodes). Simulation results show the outcome of the proposal in terms of percentage of packets losses, end-to-end average packet delay and average delay jitter.

(ii) Several works include proposals about \textit{QoS-aware multimetric routing for MANETs}. The work \cite{AQR_QoS_Video_MANET} presents an on-demand routing adaptive protocol named adaptive quality of service routing (AQR), which takes into account the QoS parameters of bandwidth, delay and cost. AQR has QoS violation detection and recovery mechanisms which enables quick re-routing of the packets along new paths. NS-2 simulations show that AQR achieves the QoS performance claimed and adapts well to different topologies and mobility conditions.
In \cite{ANFIS_QoS_Video_MANET} the authors present a modification of the basic AODV routing protocol able to create optimal routes considering the available bandwidth to fulfill the video stream requirements. The work \cite{QoS_Weng} presents a power-aware on-demand routing protocol for MANETs using bandwidth and power information gathered from the 1-hop neighbors of the nodes, in order to improve path bandwidth and reduce the overall power consumption. Similarly, in \cite{QoS_Taha} the authors introduce a fitness function technique to optimize the energy consumption in MANETs to find the optimal path from source to destination to reduce the energy consumption in multipath routing. The work \cite{QoS_Paramasivan} presents a game theoretical routing protocols for MANETs that improves QoS provision by focusing on the collaboration between nodes. Their approach minimizes the utility of malicious nodes and motivates cooperation between nodes by using a reputation system. The work \cite{QoS_Bhard} introduces a multipath routing protocol using genetic algorithm to select efficient routes that have the shortest route, maximum residual energy and less overhead. Also, we highlight our previous proposal named multipath multimedia dynamic source routing (MMDSR) \cite{Ahmad_Jitel_2015} which considers the specific characteristics of the video when it establishes a QoS-aware path to forward packets from source to destination looking to provide the best QoS. The QoS metrics considered are available bandwidth, packet losses, packet delay, packet delay jitter, hop count to destination, reliability metric and mobility metric. The MMDSR operation is detailed in section \ref{sec:QoS_routing}. 

(iii) Regarding \textit{social MANETs}, the work in \cite{integrated_social_QoS} proposes to combine \textit{social trust} metrics derived from social networks with \textit{QoS trust} metrics derived from communication networks to obtain a composite trust metric for evaluating trust of mobile nodes in MANETs. As social trust metrics they use \textit{social ties} (two nodes have a lot of direct or indirect interaction experiences with each other) and \textit{honesty} (a belief of whether a node is malicious or not). As QoS trust metrics they use \textit{energy} (residual energy of a node) and \textit{cooperativeness} (whether the node is cooperative in routing and forwarding packets). They find the best protocol settings to decrease the trust bias between subjective and objective trust evaluation results, which happens around a 80\% of using direct observations (i.e., 20\% of indirect observations from trustworthy recommenders) for subjective trust evaluation. 
In \cite{HC_social_MANET}, the authors propose a hierarchical cooperation protocol where the forwarding path from source to destination is formed taking into account the nodes' density ($\gamma$) and the number of social contacts per node (\textit{q}). They model the probability that any two nodes in distance \textit{d} away from each other are socially connected is assumed to be proportional to $d^{-\gamma}$. They analyze the throughput–delay trade-off in dense and sparse networks for different parameters (\textit{q}, \textit{d}, $\gamma$). Their results show that as $\gamma$ grows the throughput–delay trade-off improves when considering the social information between nodes.

As far as we know, there is still no proposal that addresses the issue of providing video-streaming services through MANETs using a multimetric routing protocol that includes a social metric, which is the objective of our work. In the next two sections we summarize in a nutshell the basics of our QoS-aware multimetric routing protocol MMDSR, previously presented in \cite{GT_video_MANET_Ahmad}. Afterwards, in section \ref{sec:TS} we introduce a new social-aware metric named \textit{tie strength} (TS). This TS will be included in the MMDSR forwarding scheme by means of a new QoS-aware and social-aware multimetic score, presented in section \ref{sec:mscore}.

\section{Basics of our framework to provide video-streaming services over MANETs}
\label{sec:basics}
 In this section we review the basic features of the video transmitted. Video frames are distributed using RTP/RTCP (Real-time Transport Protocol/RTP Control Protocol) over UDP as transport layer protocols.

\begin{figure}[htpb]
\centering
\includegraphics[height=0.75in, width=3in]{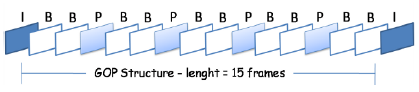}
\caption{Structure of a group of pictures (GoP) formed by IPB video frames.}
\label{fig1}
\end{figure}

The considered video-streaming service uses a basic layered MPEG-2 VBR video coder. The video flow is composed of sets of frames, usually 4 to 20 frames, named group of pictures (GoP). A GoP has three kind of video frames: I, P and B, with a frame-pattern repeated in each GoP. The composition of a GoP and the relationship between frames in the decoding process is depicted in Figure \ref{fig1}. I (Intra) frames encode spatial redundancy and form the base layer, providing a basic video quality. I frames carry the most important information for the decoder. If an I frame is lost, the whole corresponding GoP will be useless for the decoder and will be discarded. P (Predicted) and B (Bi-directional) frames carry differential information with regards to preceding (for P) or preceding and posterior (for B) frames, respectively. We considered those features to assign different to each video frame according to their importance in the video stream. Thus, I frames have the highest priority, P frames a medium priority and B frames the lowest one. 

We assume that the IEEE 802.11e \cite{IEEE} standard works in the media access control (MAC) layer. This standard provides QoS support for services such as video-streaming. The IEEE 802.11e includes four different access categories (AC). Each packet from the higher network layer arrives at the MAC layer with a given priority. That packet is classified at the MAC layer to the proper AC. We defined a mapping between the different classes of packets into one of the four ACs:

\begin{itemize}
 \item AC0: signaling
 \item AC1: high priority packets (I frames)
 \item AC2: medium priority packets (P frames)
 \item AC3: low priority packets (B frames + best effort traffic)
\end{itemize}

\section{QoS-aware multimetric routing protocol for video-streaming services over MANETs}
\label{sec:QoS_routing}
In this section we summarize the operation of our proposal called \textit{multipath multimedia dynamic source routing} (MMDSR) previously presented in \cite{Ahmad_Jitel_2015} and \cite{GT_video_MANET_Ahmad}. MMDSR is a QoS-aware multimetric routing protocol specially designed to transmit video-streaming over IEEE 802.11e \cite{IEEE} MANETs. MMDSR considers the specific characteristics of the video while establishing the forwarding path to transmit packets from source to destination, aiming to provide a good QoS.  

\subsection{Dynamic source routing (DSR)}
\label{sec:dsr}
Dynamic source routing (DSR) \cite{RFC4728} is a distance-based routing protocol where a source node establishes an end-to-end forwarding path from source to destination. Nodes periodically interchange small hello messages (also named beacons) to announce its presence to their neighboring nodes in transmission range, i.e. to their one-hop neighbors. With this information, every node builds a one-hop neighbors’ table, to be used by the routing protocol to make forwarding decisions. In DSR, the forwarding path is formed by those nodes which form the shortest path to destination. DSR is the routing engine used by our proposal MMDSR to find out all the available forwarding routes from source to destination. Then, MMDSR selects the best path to transmit the video frames according to a QoS-aware forwarding algorithm which is described in the next subsection.

\subsection{QoS-aware multimetric dynamic source routing}
\label{sec:mmdsr}
\begin{figure}[hpb]
\centering
\includegraphics[height=1.2in, width=3.2in]{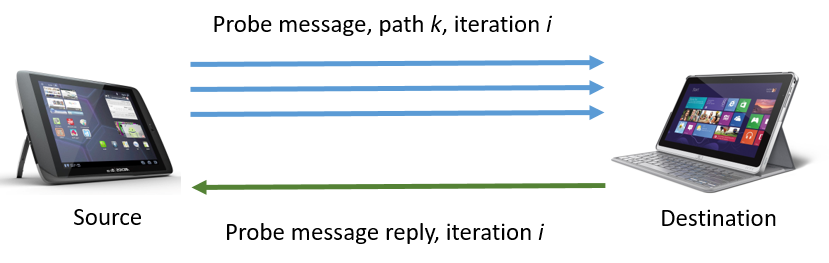}
\caption{Probe message (PM) packets periodically flooded from source to destination. Probe message reply (PMR) packets gather QoS feedback about the available paths $k$ found in iteration $i$.}
\label{fig2}
\end{figure}

Basically, MMDSR uses DSR to search all the available paths from source to destination. Over those available paths, MMDSR sends monitoring \textit{probe message} (PM) packets. Afterwards, a \textit{probe message reply} (PMR) packet generated at destination is sent back to the source. The PMR packet gathers the collected information about the quality of all the available paths from source to destination. Figure \ref{fig2} shows PM and PMR packets which are periodically interchanged between source and destination.

Those paths are then filtered to check if they fulfill the user's QoS requirements ($customer_{request}$) according to the required video quality, see Eq. (\ref{eq:customer_req}). We consider the following QoS parameters: minimum expected bandwidth ($BW_{min}$),  maximum percentage of packet losses ($L_{max}$), maximum packet delay ($D_{max}$) and maximum delay jitter ($J_{max}$).

\begin{equation}
\label{eq:customer_req}
  customer_{request} \equiv \{BW_{min}, L_{max}, D_{max}, J_{max}\}
\end{equation}

The paths that fulfill the user's requirements are then arranged according to a QoS multimetric score assigned to each one of the paths using the feedback QoS information carried in the PMR packets received at the source node.

\begin{equation}
\label{eq:path_state}
  QoS\_multimetric\_score_k^i \equiv \{BW, L, D, J, H, RM, MM\}_k^i
\end{equation}

\noindent
where \textit{i} is the algorithm iteration and \textit{\textit{k}} means each one of the available paths. The QoS metrics used to compute the multimetric score associated to each path $k$, are: end-to-end available bandwidth ($BW_k^i$), percentage of packet losses ($L_k^i$), average end-to-end packet delay ($D_k^i$), average packet delay jitter ($J_k^i$), hop-count distance ($H_k^i$), reliability metric ($RM_k^i$) calculated from the signal-to-noise ratio (SNR) of the links involved in each path $k$, and mobility metric ($MM_k^i$) calculated from the relative mobility of the neighboring nodes within each
path $k$.

The set of available paths from source to destination is updated periodically (iteration $i$ by iteration) because of the dynamic topology of MANETs, whose nodes move producing frequent link breakages. Finally, the source node selects the best path (i.e., the one with the highest score) over which the video frames will be transmitted.  

\subsubsection{Dynamic self-configuration of the video-streaming framework}
\label{sec:self_conf}
In this section we summarize the basics of the self-configuration operation, which is fully detailed in \cite{GT_video_MANET_Ahmad}. Due to the inherently dynamic network topology of MANETs, suitable routing protocols should be dynamically self-configured.

In this sense, MMDSR periodically monitors the current state of the MANET. Upon detection of network changes, the routing protocol modifies the refreshing period $T_{routing}$ to update the set of available paths from source to destination. To do so, we use a tuning function (expressed below in Eq. (\ref{eq:Trouting})) to dynamically adjust the refreshing period  depending on a parameter called \textit{NState}. \textit{NState} gathers information about the global network state at each iteration $i$, as follows:

\begin{equation}
\label{eq:Nstate}
 \begin{split}
  NState^i & = w_{RM} . \overline{RM^i} + w_{MM} . \overline{MM^i} + w_{BW} . \overline{BW^i} + w_{L} . \overline{L^i}\\
               &\quad + w_{D} . \overline{D^i} + w_{J} . \overline{J^i} + w_{H} . \overline{H^i}
 \end{split}
\end{equation}

In the previous equation, the upper bars mean the average for each QoS parameter calculated for each available path $k$. The weights $w_i$ have suitable values that sum one. In this work we have set them equally to $1/7$, although other values could be chosen to highlight a metric over the others in the computation of the network state. When a source receives a new PMR packet with feedback information from the network  (algorithm iteration $i$), it updates the $NState^i$ with Eq. (\ref{eq:Nstate}).

Finally, we design a proper tuning function to dynamically update the period $T_{routing}$ to refresh the set of available paths from which the best path is selected:

\begin{equation}
\label{eq:Trouting}
  T_{routing}^{i+1} = \alpha \cdot NState^i + \beta
\end{equation}

To obtain the previous equation, a high number of simulations were conducted under a wide range of network conditions where the network performance was \textit{good}, \textit{normal} or \textit{bad}. For the scenario under consideration used in the present work, the obtained values were 
$\alpha$ = 10 and $\beta$ = 3. The idea behind this equation is that when the global network state is in good conditions (i.e., $NState^i$ is high) the current forwarding paths can be used longer (i.e., $T_{routing}$ increases), whereas when the network state is in bad conditions (i.e., $NState^i$ is low) the forwarding scheme should be updated sooner (i.e., $T_{routing}$ decreases).

\section{Tie strength applied to MANETs}
\label{sec:TS}
The idea of measuring relationship strength among users in a MANET scenario will allow us to select the path to forward packets from source to destination according to a forwarding algorithm designed not only based on QoS parameters, but also on a certain level of confidence between the forwarding nodes from source to destination. Our routing proposal is designed to balance the trade-off in the forwarding path between confidence level and QoS parameters (percentage of packet losses and average end-to-end packet delay). To attain or goal, in Section \ref{sec:ts_computation} we define a new social metric named \textit{tie strengh} to be included in the computation of the multimetric score to classify the available forwarding paths. 

It is important to notice that in social media there is no distinction between users, i.e. they are treated all the same either as trusted friend or stranger. This topic has been investigated for years and it is somehow assessed with the \textit{tie strength} concept. The idea of closeness between users in social media has been analyzed in different application fields, such as recommendation of goods \cite{song2017whose}, the word-of-mouth spread of information \cite{choi2017wom}, or even has been explored to know how the closeness between two users can be extended to a triadic (group of three people) \cite{huang2018will}. Our work is inspired by a previous research work developed in \cite{Servia13}, where they introduced a predictive model that maps social media data to tie strength concept. This approach was also used in \cite{servia2015evolution} to assess how the co-authorship network is related to the research performance over time. In this section, we will summarize briefly some of the basic ideas built up to be applied in this new field of study.

\subsection{Social spheres}

A \textit{social sphere} represents the social contacts and topics of interest of a user when using social networks. In a previous research work \cite{Servia13}, a model was built exclusively taking into account the interaction data gathered from public APIs (with users' permission) in Facebook and Twitter. Interesting interaction evidences or tie signs (private messages, retweets, mentions, etc.) were firstly identified, as it is summarized in Table \ref{tab:fSigns} (Facebook evidences) and Table \ref{tab:tSigns} (Twitter evidences). Then, they were analyzed to assess the strength of the social ties between users. For this to be possible, tie signs were tagged as direct or indirect interactions and as public or private interactions. In order to contextualize these tie signs, different social contexts were also taken into account, such as personal, familiar, professional, etc. Therefore, the tie signs between two users are weighted according to the following parameters: the different categories of the tie sign or evidence (private/public, direct/indirect) and the different context in which the tie sign occurred. 

\begin{table*}[htpb]
\centering
  \caption{Tie signs classification on Facebook \cite{Servia13}.}
  \label{tab:fSigns}
  \begin{tabular}{lcccc}

    Tie signs ($S_{u|k}(v)$)  &Direct&Indirect&Public&Private\\

    Wall-posts in friend's Wall&x& & &x\\
    Private messages exchanged&x& & &x\\        
    Comments in friend's objects&x& &x& \\
    Comments in the same objects& &x&x& \\
    Likes in friend's objects&x& &x& \\
    Likes in the same objects& &x&x& \\
    Being tagged in the same photos or videos& &x& &x\\
    Belonging to the same private group& &x& &x\\
    Belonging to the same public group& &x&x& \\
    Attending to the same private event& &x& &x\\
    Attending to the same public event& &x&x& \\
    Being subscribed to the same user& &x&x& \\
    Being subscribed by the same user& &x&x& \\

\end{tabular}
\end{table*}

\begin{table*}
\centering
  \caption{Tie signs ($S_{u|k}(v)$) classification on Twitter \cite{Servia13}.}
  \label{tab:tSigns}
  \begin{tabular}{lcccc}

    Tie signs ($S_{u|k}(v)$)  &Direct&Indirect&Public&Private\\
 
    Mentions (replies)&x& & &x\\
    Private messages exchanged&x& & &x\\        
    Retweets friend's tweets&x& &x& \\
    Retweets the same tweets& &x&x& \\
    Marking as favorite friend's tweets&x& &x& \\
    Marking as favorite the same tweets& &x&x& \\
    Taking part of the same private list& &x& &x\\
    Taking part of the same public list& &x&x& \\
    Sharing the same Hashtag& &x&x& \\
    Common followers& &x&x& \\
    Common followees& &x&x& \\    

\end{tabular}
\end{table*}

There is a web service, coined as \textit{mySocialSphere}, that can be accessed throughout a public API to obtain the public social sphere of any user. Thus, given a user $u$, this web service analyzes the strength of his/her tie signs with other users and provides a tie strength index (a value between 0 and 1, being 1 the highest tie strength possible) that assesses the number and quality of the interactions of $u$ with any other user $v$. It is needed to remark that the tie strength index is inherently asymmetric, since the tie strength perceived for $u$ about his/her relationship with $v$ may be different from the tie strength perceived for $v$ about his/her relationship with $u$. A simple and extreme example is the following one: user $u$ often chats with $v$ but also with other users, whereas $v$ only chats with $u$, the tie from $v$’s perspective will be stronger than from $u$’s one.

{Although all the mathematical details to compute the tie strength between two users are explained in \cite{Servia13}, we consider that it is adequate for the purpose of this paper to mention here the following most relevant aspects. The tie strength of the relationship between user $u$ and user $v$, from the user $u$'s perspective, is denoted as $T_{u}(v)$ and it is obtained as follows}:

\begin{equation}
\label{eq:Tuv}
    T_{u}(v) = \sum_{k=1}^{N} \alpha_{k} \cdot f (|S_{u|k}(v)|), \mbox{ \; \text{where} \; }  \sum_{k=1}^{N} \alpha_{k} = 1
\end{equation}

\noindent
{where $S_{u|k}(v)$ refers to the number of interactions of type $k$ that $u$ has with $v$; $N$ is the number of different tie signs analyzed (shown in Table \ref{tab:fSigns} and Table \ref{tab:tSigns}); and $k$ refers to the type of interaction. $\alpha_{k}$ denotes the weights of an interaction that belongs to type $k$ (i.e., the influence of each type of interaction in the calculation of the tie strength). For instance, private interactions have a higher weight than public interactions; consequently, the $\alpha_{k}$ is higher for private interactions than for public ones. Finally, function $f$ is the following normalisation function:} 

\begin{equation}
{
    f(x) = 
    \begin{cases} 
        0 & \mbox{if $0 \leq x \leq \frac{\overline{x}^{2}}{x_{max}}$} \\
        \frac{\ln{\frac{x_{max} \cdot x }{\overline{x}^{2}}}}{\ln{\frac{x^{2}_{max} }{\overline{x}^{2}}}}  & \mbox{if $\frac{\overline{x}^{2}}{x_{max}} < x$}
    \end{cases} 
    }
\end{equation}

\noindent
{where $\overline{x}$ is the mean and $x_{max}$ the maximum value of variable $x$. Therefore, $f(x)$ tends to $1$ if $x > \overline{x}$, $f(x)$ tends to $0$ if $x < \overline{x}$ and $f(x)$ tends to $0.5$ if $x \sim \overline{x}$. Function $f(x)$ guarantees that $0 \leq T_{u}(v) \leq 1$.}

Finally, tie strength indexes are dynamic values that must be constantly updated (i) to take into account new interaction evidences and give them more relevance in the index calculation and (ii) to reflect that old interaction evidences are progressively less important and, consequently, have less relevance in the index calculation \cite{Servia13}.  {With this aim, we apply a decreasing exponential function to Eq. (\ref{eq:Tuv}) as follows:}

\begin{equation}
    T_{u}(v, t) = e^{-\mu \cdot textit{t}} \cdot T_{u}(v)
\end{equation}

\noindent
{where $\mu$ is the decreasing factor and $t$ is the time since the latest updated of the sign.} Notice that, in order to obtain a representative value of the tie strength that for $u$ has his/her relationship with $v$, the index has to be obtained when needed, since historic values may not properly reflect the current interaction between them.

\subsection{Computation of average tie strength of a forwarding path in a MANET}
\label{sec:ts_computation}
We define five integer values to measure the strength of a tie (\textit{tie strength}) between an individual and each one of the other individuals in a MANET. This way, TS takes integer values from 0 to 4, where 4 represents the strongest tie (high interaction and/or trust) and 0 represents the weakest tie (very poor or no interaction and/or untrusted relationship). We have chosen five values following the same structure traditionally done in the mean opinion score (MOS) metric used to assess subjective quality of service (QoS) in multimedia applications \cite{MOS}. Notice that the tie strength (TS) is asymmetrical (i.e., $TS_u(v) \neq TS_v(u)$ ) and that the tie between a given user and itself is 0 (i.e., $TS_u(u) = 0$). 

\begin{figure}[htpb]
\centering
\includegraphics[height=2.5in, width=3.6in]{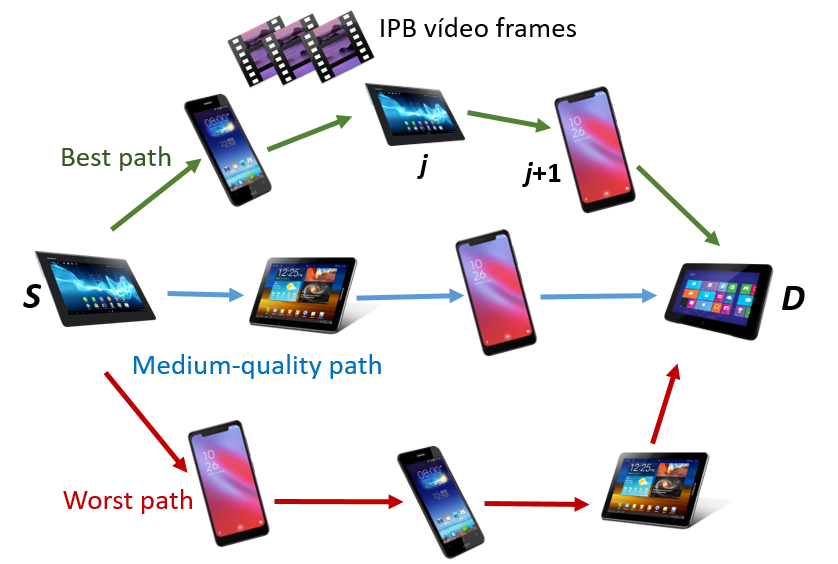}
\caption{General MANET scheme to transmit video frames from a source $S$ to its destination $D$. Paths are arranged according to a combined QoS and social multimetric score, see Eq. (\ref{eq:multimetric_score}).}
\label{fig:MANET}
\end{figure}

Finally, we define the average tie strength $\overline{TS_k}$ of a path $k$ by means of the geometric mean of each $TS_{k, j}$ between every node $j$ of the path and its neighbor $j+1$, from a source $S$ to its destination $D$, see Eq. (\ref{eq:TS}), for a MANET general scheme represented in Figure \ref{fig:MANET}. $NF_k$ stands for the number of forwarding nodes that compose path $k$.

\begin{equation}
\label{eq:TS}
  \overline{TS_k} = \sqrt[NF_k]{\prod_{j=1}^{NF_k}TS_{k, j}(j+1)} \hspace{2 pt}  , \hspace{6 pt} 0 \leq TS_{k, j} \leq 4
\end{equation}

\section{Algorithm to arrange the available paths}
\label{sec:mscore}

In this section we present our proposal of algorithm to arrange the available paths found from source to destination. Finally, our \textit{QoS-aware social-aware routing protocol to provide video-streaming services over MANETs} (QSMVM), will select the forwarding path with the highest multimetric score, see Eq. (\ref{eq:multimetric_score}).

Once the source has filtered out the set of paths that meets the user's QoS requirements (see Eq. (\ref{eq:customer_req})), a multimetric score is computed for every path, taking into account the qualifications of the QoS parameters as well as the tie strength (TS) qualification. This multimetric score is computed as follows, for each path $k$ at each iteration $i$ of the algorithm:

\begin{equation}
\label{eq:multimetric_score}
  MScore_k^i = w_{QoS} \cdot (BW + L + D + J + H + RM + MM)_k^i + w_{TS} \cdot \overline{TS_k^i}
\end{equation}

\noindent
{where} $w_{QoS}$ and $w_{TS}$ are weights that sum one, see Eq. (\ref{eq:weigths}). In Eq. (\ref{eq:multimetric_score}), $w_{QoS}$ is the weight given to the set of QoS parameters. In this work we give the same weight to all QoS metrics. Alternatively, we could also assign different weights if we want some parameter to have more importance than the others in the $MScore_k^i$ calculation. Besides, in Eq. (\ref{eq:multimetric_score}) $w_{TS}$ is the weight given to the TS of path $k$. The computation of the TS associated to each path is computed with Eq. (\ref{eq:TS}). 

\begin{equation}
\label{eq:weigths}
  w_{QoS} = 1-w_{TS}
\end{equation}

Once computed the $MScore_k^i$ for each available path $k$ in iteration $i$, the source arranges them from the highest-quality path to the worse-quality path. Finally, the source selects the path with the highest $MScore_k^i$ score.
Notice that as the  $w_{TS}$ increase, the average level of confidence among nodes forming a particular path becomes more decisive in the $MScore$ and therefore, in the selection of the best path. Nonetheless, there will be a trade-off between the confidence level among the forwarding nodes that compose the selected path and the QoS provided by that path. 

In the next section we will carry out a performance evaluation to asses the benefits of our proposal regarding the QoS in terms of packet losses and packet delay, and also in terms of confidence level of the relying nodes. Additionally, we will analyse the trade-off between QoS and confidence level of the forwarding path.

\section{Performance evaluation and simulation results}
\label{sec:simulations}
We have used the open source network simulator ns-2 (v2.27) \cite{ns2} to implement and evaluate the performance of our proposal named \textit{QoS-aware and social-aware multimetric routing protocol for video-streaming services over MANETs} (QSMVM). We have carried out a set of simulations with 5 repetitions per each simulation point with independent scenario seeds and all the figures show confidence intervals of 90\%. We have used the Bonnmotion tool \cite{bonmotion} to generate the MANET scenarios. Furthermore, interfering CBR traffic was generated to constrain the paths. There are two source nodes and two destination nodes, so that each source sends packets related to a video-streaming service. The evaluation is done under two different nodes' densities: (i) a medium nodes' density (100 $nodes/km^2$) and (ii) a high nodes' density (200 $nodes/km^2$). Simulation settings of the scenarios are shown in Table~\ref{tab:simSett}.

We analyse the effect of $w_{TS}$ regarding the value of the multimetric score $MScore^i_k$ in path $k$ for iteration $i$ (see Eq. (\ref{eq:multimetric_score})). We make $w_{TS}$ vary from 0 (i.e., only QoS parameters are considered) to 1 (only the average TS is considered) and see the impact of giving more importance to the level of confidence of a path ($\overline{TS}$) over the QoS metrics. Specifically, we analyse  the effect in the percentage of packet losses and in the average end-to-end packet delay. 

We have generated different scenarios regarding the social confidence level between couples of users. Accordingly, we have scenarios from low tie strength values between every couple of users (i.e., a scenario with low social media interaction between users), to scenarios with high tie strength values between couple of users (i.e., a scenario with friends and acquaintances, with a high social media interaction). To obtain the diverse scenarios we generate values of tie strength (TS) between nodes following a normal distribution with a mean taken values from $\mu=1$ to $\mu=4$, with standard deviation of $\sigma=1$ for all the scenarios. The choice of these parameters gives us a very wide range of possible TS values between pairs of individuals in the population considered. As a result, we will obtain a large representative number of different simulation scenarios. 

Notice that since TS is an asymmetrical measure, TS from a given user $u$ towards a user $v$ is different to the TS from user $v$ towards user $u$ ($TS_{u-v} \neq TS_{v-u}$). Also, notice that TS between a given mobile node and itself is zero ($TS_{u-u}=0$). The simulation settings of the evaluated scenarios are shown in Table~\ref{tab:simSett}. The scenarios used to test the proposal consists of a set of 27 and 54 mobile nodes distributed in a MANET of 520 m x 520 m, producing nodes' densities of 100 and for 200 $nodes/km^2$, respectively. The transmission range of the nodes is 120 m. Nodes move with a speed up to 2 m/s. Two video flows are transmitted from node $S_1$ to $D_1$ and from node $S_2$ to $D_2$, respectively.

\begin{table}
\centering
  \caption{Simulation settings of the scenarios evaluated.}
  \label{tab:simSett}
  \begin{tabular}{lc}

    Area&520 m x 520 m \\        
    Number of nodes &27 and 54\\
    Nodes' density &100 and 200 nodes/km$^2$\\
    Average node speed&2 m/s \\
    Transmission range&120 m \\
    Mobility Pattern&Random Waypoint \\
    MAC specification&IEEE 802.11e, EDCF \\
    Nominal bandwidth&11 Mbps \\
    Simulation time&200 s \\
    Video codification&MPEG-2 VBR \\
    Video bit rate&150 Kbps \\
    Number of video sources& 1 \\
    Video streamed&Blade Runner \\
    MANET routing protocol & QSMVM, see Sec. \ref{sec:mscore} \\
    Transport layer protocols&RTP/RTCP/UDP \\
    Maximum packet size&1500 Bytes \\
    QoS weighting values (See Eq. (\ref{eq:Nstate}))&1/7 \\
    Tie strength (TS) normal distribution & $\mu=1$ to $\mu=4$, $\sigma^2=1$ \\
    Queue sizes&50 packets \\
    Interfering CBR traffic&300 Kbps \\
    Channel noise -92 dBm&-92 dBm \\
    Mobility generator&Bonnmotion \cite{bonmotion} \\

\end{tabular}
\end{table}

Figures \ref{fig:100_mu1_losses_TS} to \ref{fig:200_mu4_delay}  show the average percentage of packet losses and the average end-to-end packet delay, for a TS normal distribution whose parameters are $\sigma=1$ and $\mu$ = 1 to $\mu$ = 4. {Notice that the percentage of packet losses, which show values around $20\%$, apparently might seem quite high. However, we must contextualize those results to the inherent low-connectivity feature of MANETs. MANET nodes constantly move, which produces frequent link breakages and consequently packet losses increase. To cope with this, our multimetric proposal MMDSR notably outperforms vanilla distance-based DSR, showing an improvement around 10 dB in the PSNR \cite{QoS_video1}}.

We compare the results to the case of not considering the average tie strength $\overline{TS}$ ($w_{TS}=0$) in the path selection process, i.e. we always select the path with best QoS performance without taking the level of confidence among forwarding nodes into account.

In the x-axis we vary the weight $w_{TS}$ in Eq. (\ref{eq:multimetric_score}) to give more or less importance to the social metric in front of the traditional QoS metrics. Therefore, in the computation of the multimetric score $MScore$ used to arrange the available forwarding paths, see Eq. (\ref{eq:multimetric_score}) we have changed the weight $w_{TS}$ as follows:

\begin{itemize}
    \item When $w_{TS}$=0 we do not consider the social TS metric in the computation of the multimetric score $MScore$. That is, the forwarding path selected is the one with the highest QoS performance no matter the level of confidence among relying users.
    \item When $w_{TS}$=0.125, we have $w_{QoS}$=0.875, see Eq. (\ref{eq:weigths}). Since there are 7 QoS metrics (see Eq. (\ref{eq:multimetric_score})), in this case all metrics (i.e., each QoS metric and the TS metric) have the same weight in the computation of $MScore$. That is, each individual QoS metric would have a weight of 0.875/7=0.125, which equals $w_{TS}$.
   \item When $w_{TS}$=1, we only consider the social TS metric in the computation of the multimetric score $MScore$ and we do not consider any QoS metric. 
   \item We also consider intermediate values of $w_{TS}$, ranging from 0.2 to 0.8.
\end{itemize}

\subsection{Results for a very low social media interaction. TS normal distribution with mean $\mu=1$, standard deviation $\sigma=1$.}
 
In this set of scenarios, according to Figures  \ref{fig:100_mu1_losses_TS} to \ref{fig:200_mu1_delay} we can notice a low variation of values as $w_{TS}$ grows. This has sense because although we increase $w_{TS}$, the qualification of the $\overline{TS_k}$ parameter of the paths $k$ (see Eq. (\ref{eq:TS})) most likely will be very low (near 0) since we are in a scenario with a very low tie strength between users. This means that QoS parameters outweigh TS in the computation of $MScore$, see Eq. (\ref{eq:multimetric_score})).

\subsubsection{Very low social media interaction. Node's density = 100 nodes/km$^2$}
\begin{figure}
\centering
\includegraphics[height=2in, width=3in]{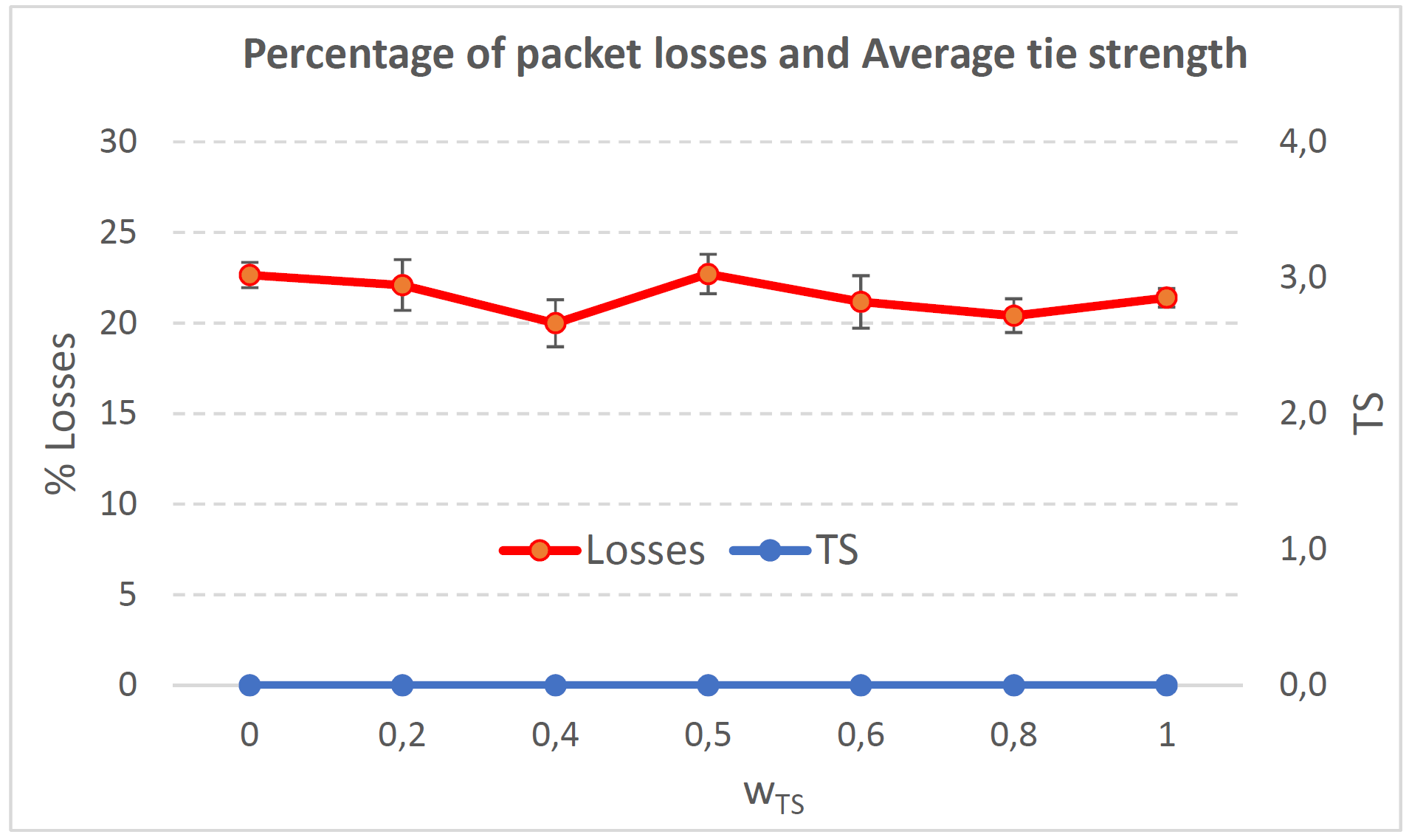}
\caption{Percentage of packet losses and average tie strength of the forwarding path. Scenario with very low social media interaction (TS normal distribution with $\mu=1$ and $\sigma=1$). Nodes' density = 100 nodes/km$^2$.}
\label{fig:100_mu1_losses_TS}
\end{figure}

\begin{figure}
\centering
\includegraphics[height=2in, width=3in]{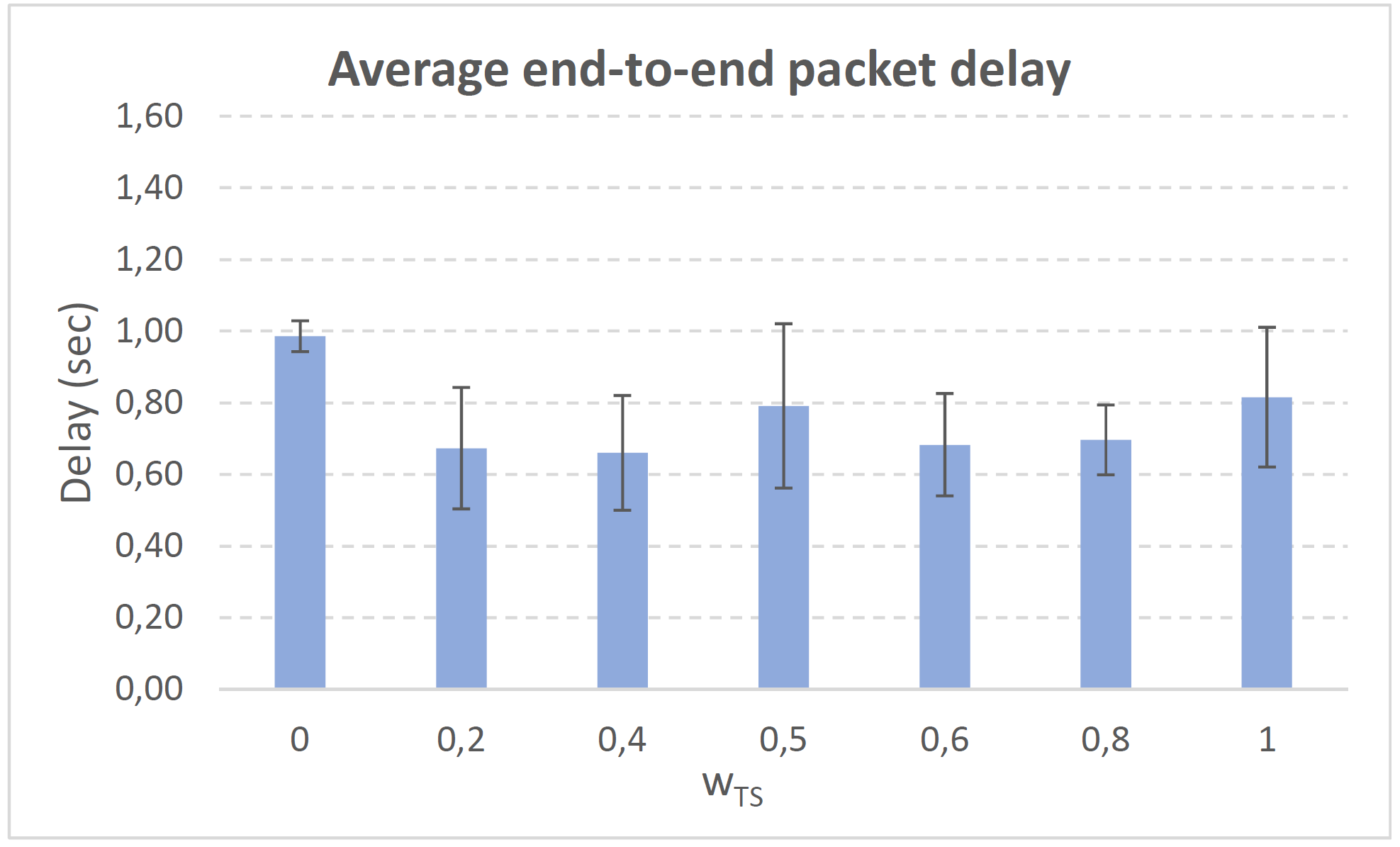}
\caption{Average end-to-end packet delay. Scenario with very low social media interaction (TS normal distribution with $\mu=1$ and $\sigma=1$). Nodes' density = 100 nodes/km$^2$.}
\label{fig:100_mu1_delay}
\end{figure}

We can see in Figure \ref{fig:100_mu1_losses_TS} that the values for the percentage of packet losses  are around 21$\%$, for all values of $w_{TS}$. There is no much variation in packet losses when we vary the weight $w_{TS}$ in the computation of the multimetric score to arrange the available paths, see Eq. (\ref{eq:multimetric_score}). This has sense because although we increase $w_{TS}$, the qualification of the $\overline{TS_k}$ parameter of the paths (see Eq. (\ref{eq:TS})) will be very low since we are in a scenario with very low tie strength between users. This means that only QoS parameters have importance in the computation of $MScore$, see Eq. (\ref{eq:multimetric_score})).  Actually, the $\overline{TS_k}$ measured in the paths $k$ taken by packets from source to destination (see blue line in Fig. (\ref{fig:100_mu1_losses_TS})), for all the cases was always 0 in all the simulations. Regarding the average end-to-end packet delay, it is around 0.7 sec for all $w_{TS}$ values. 

To sum up, in this scenario with very low social media interaction among users, the forwarding paths were formed by nodes with no tie strength between pairs of nodes. Thus, this scenario will be a reference to compare the rest of the scenarios where the social media interaction among users increases.

\subsubsection{Very low social media interaction. Node's density = 200 nodes/km$^2$}
\begin{figure}
\centering
\includegraphics[height=2in, width=3in]{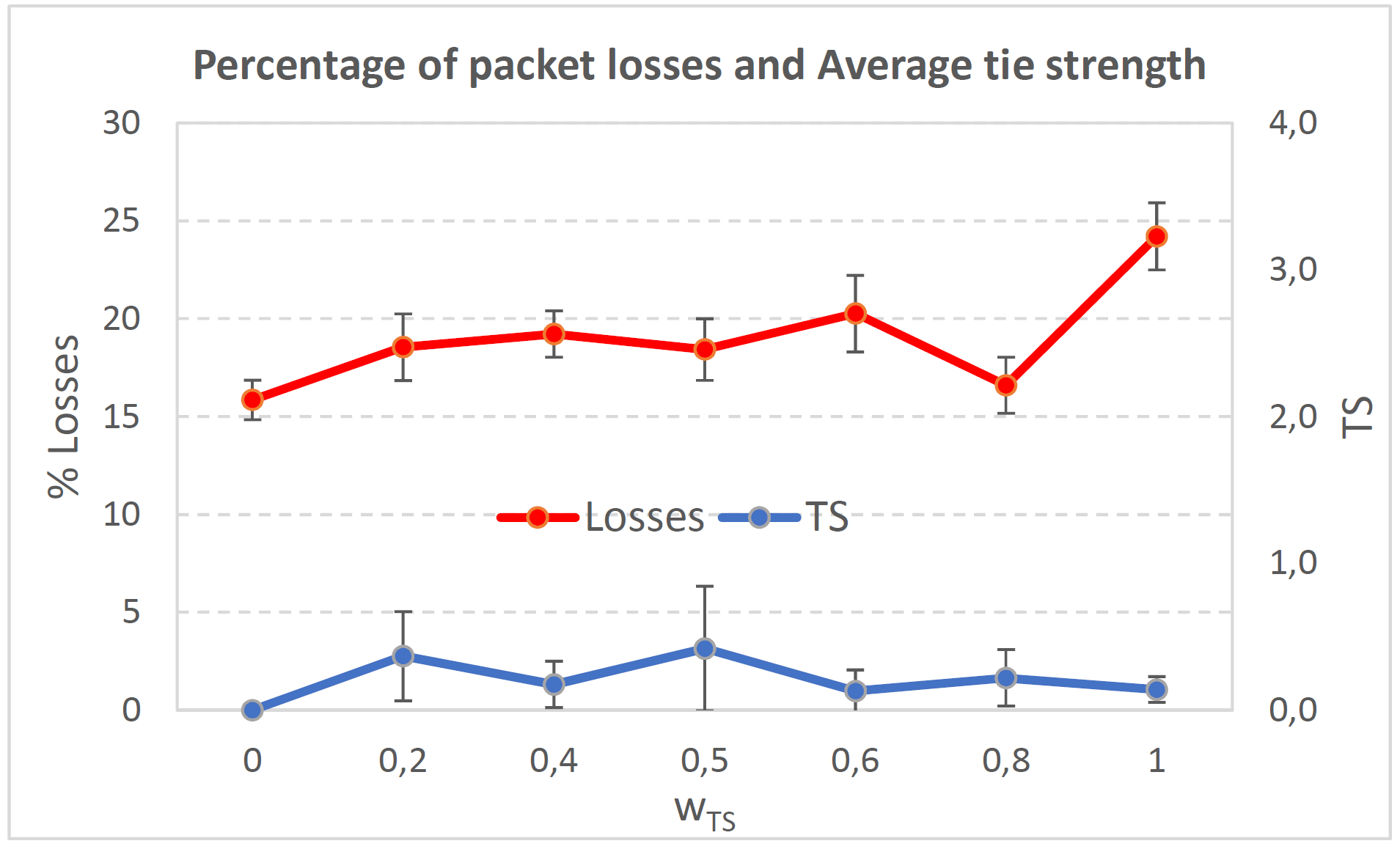}
\caption{Percentage of packet losses and average tie strength of the forwarding path. Scenario with very low social media interaction (TS normal distribution with $\mu=1$ and $\sigma=1$). Nodes' density = 200 nodes/km$^2$.}
\label{fig:200_mu1_losses_TS}
\end{figure}

\begin{figure}
\centering
\includegraphics[height=2in, width=3in]{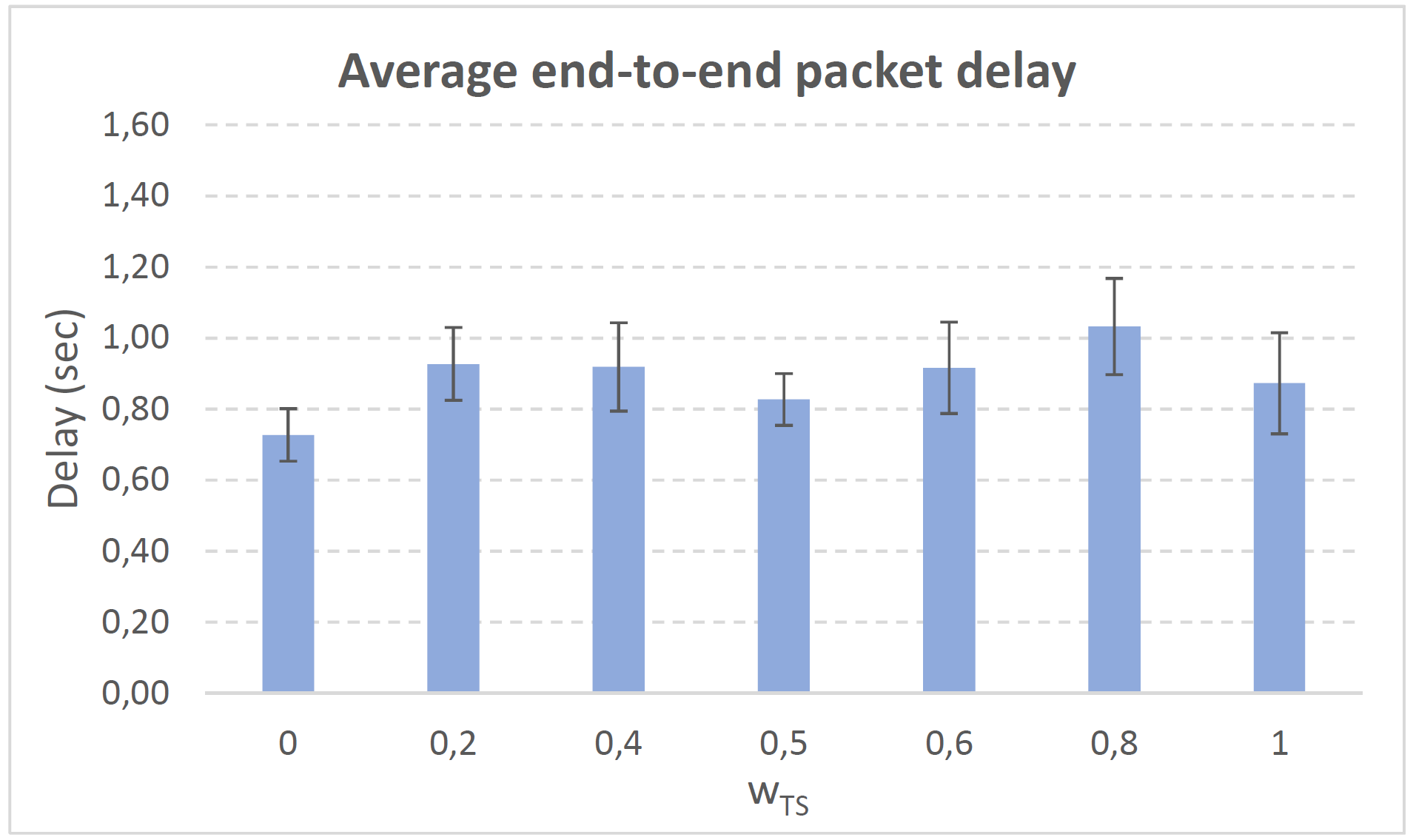}
\caption{Average end-to-end packet delay. Scenario with very low social media interaction (TS normal distribution with $\mu=1$ and $\sigma=1$). Nodes' density = 200 nodes/km$^2$.}
\label{fig:200_mu1_delay}
\end{figure}

In this scenario with a higher nodes' density with very low social interaction, the forwarding paths established from source to destination shows a variable $\overline{TS}$ around 0.3, measured in the simulations when we vary $w_{TS}$ from 0.2 to 0.8, see the blue line in Fig. (\ref{fig:200_mu1_losses_TS}). For $w_{TS}=0$ (i.e., $w_{QoS}=1$ according to Eq. (\ref{eq:weigths})), the selection of the best forwarding path is done considering only QoS parameters, and consequently the percentage of packet losses is low, around $15\%$ (see red line in Fig. (\ref{fig:200_mu1_losses_TS})). For $w_{TS}=1$ (i.e., $w_{QoS}=0$), the selection of the best forwarding path is done considering only the social TS parameter, and consequently the percentage of packet losses is the highest, around $25\%$.

The average end-to-end packet delay is around 0.9 sec for all $w_{TS}$ values, slightly higher than for the scenario with 100 nodes/km$^2$. Also, the percentage of packet losses is slightly lower in this case, around $19\%$, see Fig. (\ref{fig:200_mu1_losses_TS}). The reason is that for a higher nodes' density of 200 nodes/km$^2$ the network connectivity improves compared to the case of 100 nodes/km$^2$, and consequently packet losses decreases. Thus, packet delay increases as more packets that took longer routes made it to their destination.

\subsection{Results for a low social media interaction. TS normal distribution with mean $\mu=2$, standard deviation $\sigma=1$.}

For a scenario with a low social media interaction between users, the tie strength (TS) between consecutive nodes in the forwarding nodes from source to destination is going to increase its importance in the forwarding decision and will produce higher values in the $\overline{TS}$ measured in the simulation.

\subsubsection{Low social media interaction. Node's density = 100 nodes/km$^2$}

\begin{figure}
\centering
\includegraphics[height=2in, width=3in]{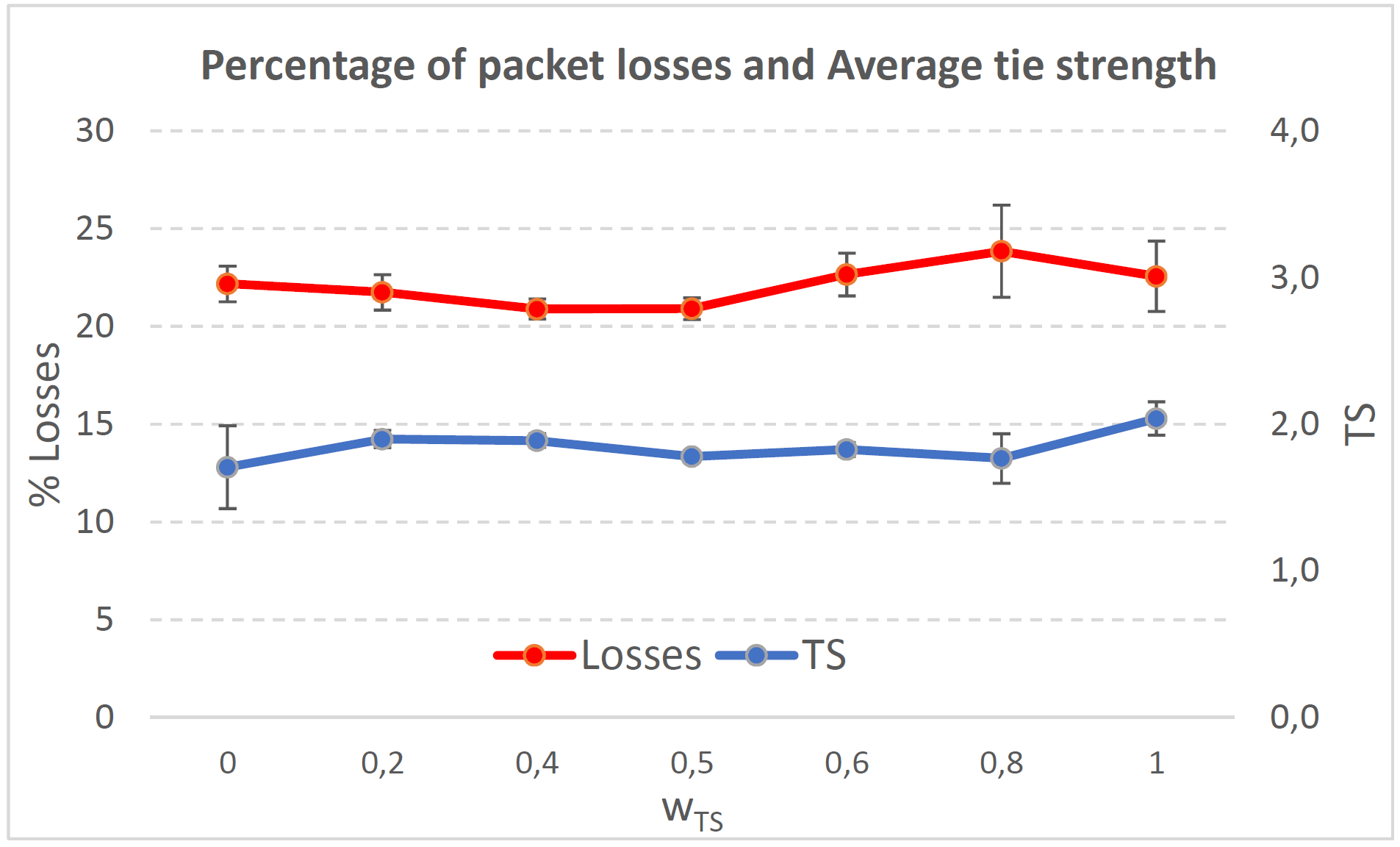}
\caption{Percentage of packet losses and average tie strength of the forwarding path. Scenario with low social media interaction (TS normal distribution with $\mu=2$ and $\sigma=1$). Nodes' density = 100 nodes/km$^2$.}
\label{fig:100_mu2_losses_TS}
\end{figure}

\begin{figure}
\centering
\includegraphics[height=2in, width=3in]{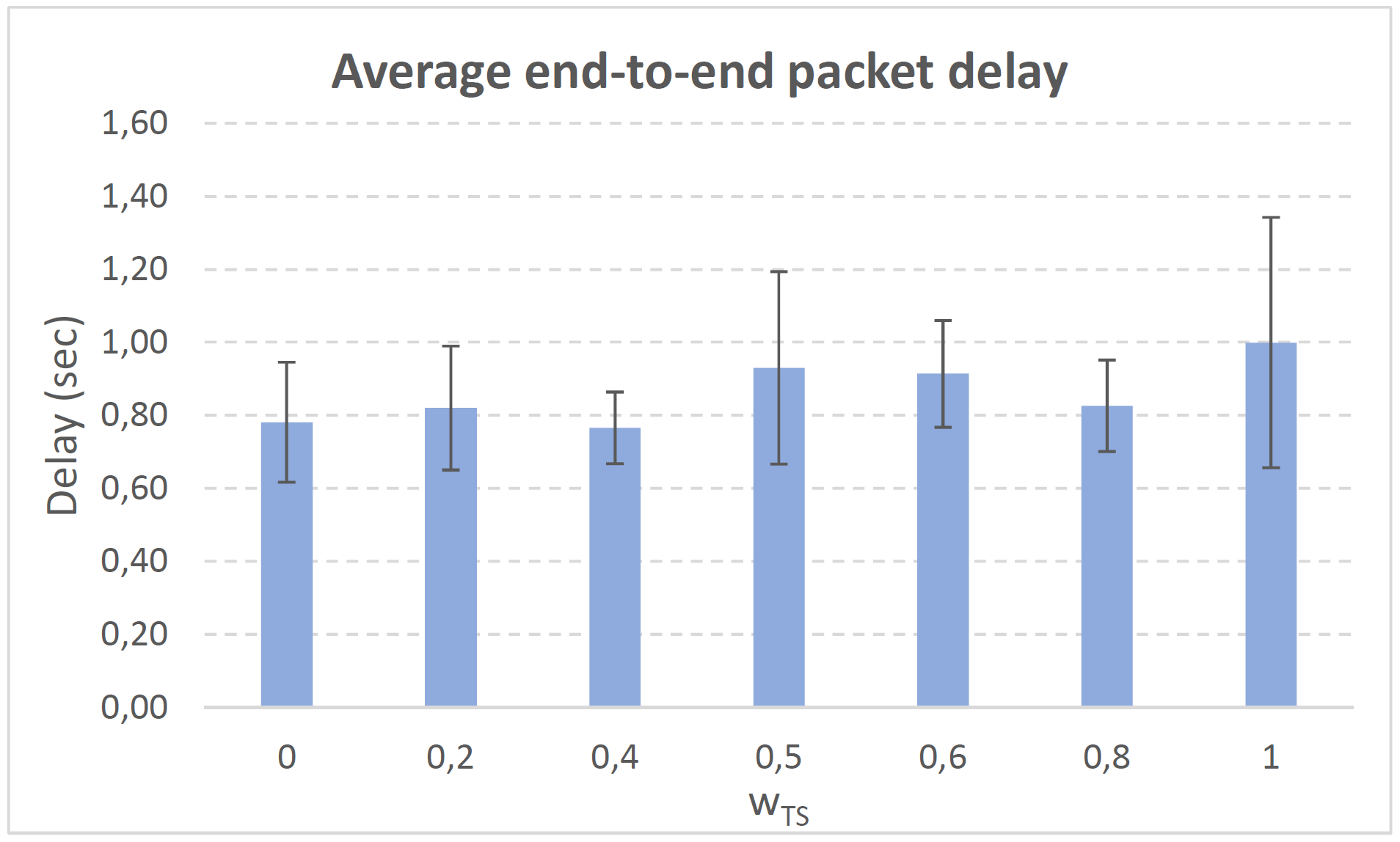}
\caption{Average end-to-end packet delay. Scenario with low social media interaction (TS normal distribution with $\mu=2$ and $\sigma=1$). Nodes' density = 100 nodes/km$^2$.}
\label{fig:100_mu2_delay}
\end{figure}

In Fig. \ref{fig:100_mu2_losses_TS} we can see a percentage of packet losses between 20\% and 24\% for $0 \leq w_{TS} \leq 1$. Again, there is no much variation in the packet losses when we vary the weight $w_{TS}$, since the TS values between pairs of neighbors are low. Consequently, the measured $\overline{TS}$ of the established forwarding paths is low and quite stable for any $w_{TS}$ (around 1.8, see blue line in Fig. \ref{fig:100_mu2_losses_TS}). A suitable value for $w_{TS}$ could be 0.4, with a 21\% of packet losses, an average confidence measured $\overline{TS}$=1.8 and an average packet delay of 0.76 sec.

\subsubsection{Low social media interaction. Node's density = 200 nodes/km$^2$}

\begin{figure}
\centering
\includegraphics[height=2in, width=3in]{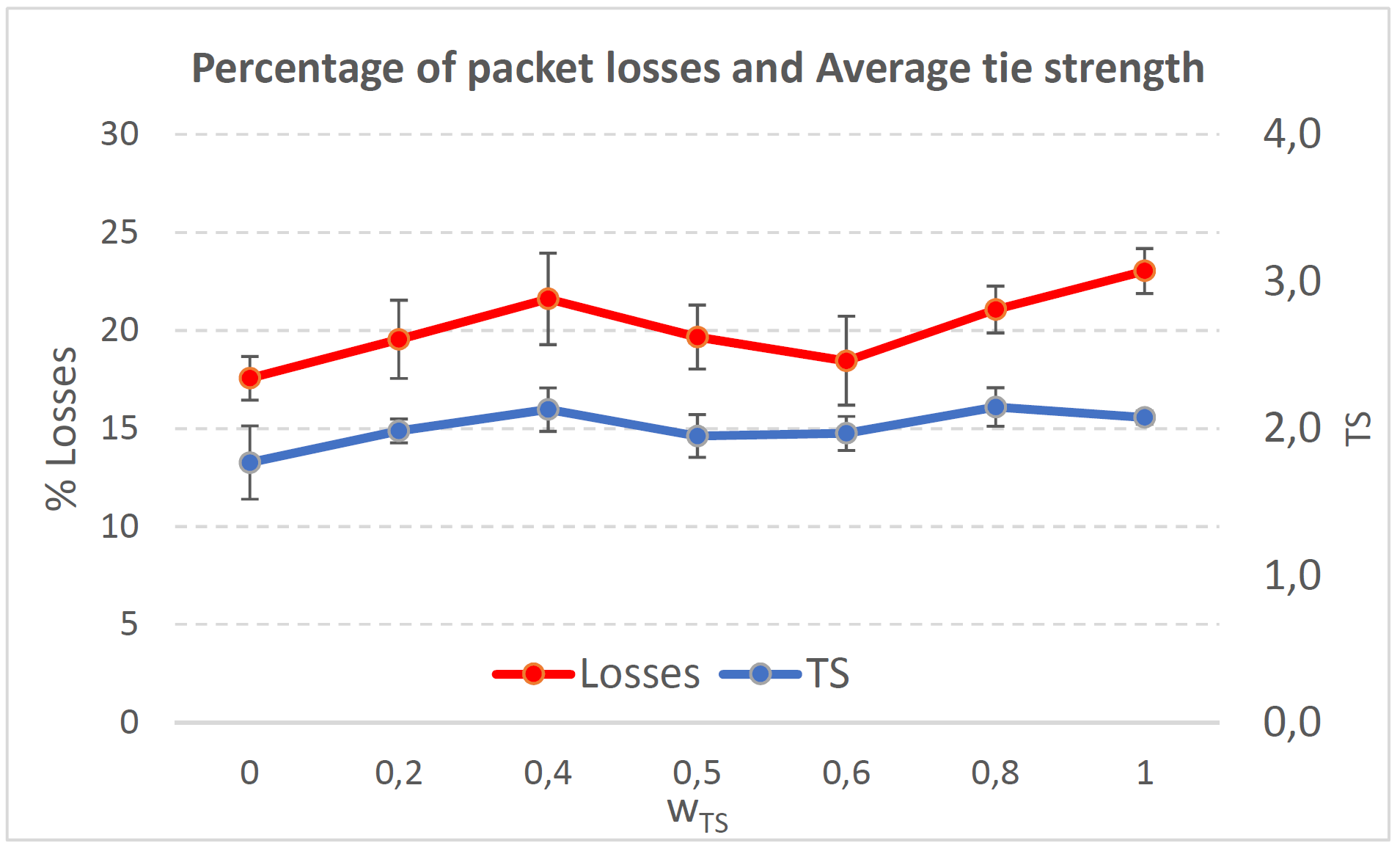}
\caption{Percentage of packet losses and average tie strength of the forwarding path. Scenario with low social media interaction (TS normal distribution with $\mu=2$ and $\sigma=1$). Nodes' density = 200 nodes/km$^2$.}
\label{fig:200_mu2_losses_TS}
\end{figure}

As it happened in the previous case with a very low social media interaction, here as we increase the nodes' density the network connectivity increases and there is a higher chance to establish a forwarding path with nodes whose confidence level is higher. Therefore, as $w_{TS}$ increases (we give more importance to the confidence level than to the QoS) the measured $\overline{TS}$ will increase (see blue line in Fig. \ref{fig:200_mu2_losses_TS}).

\begin{figure}
\centering
\includegraphics[height=2in, width=3in]{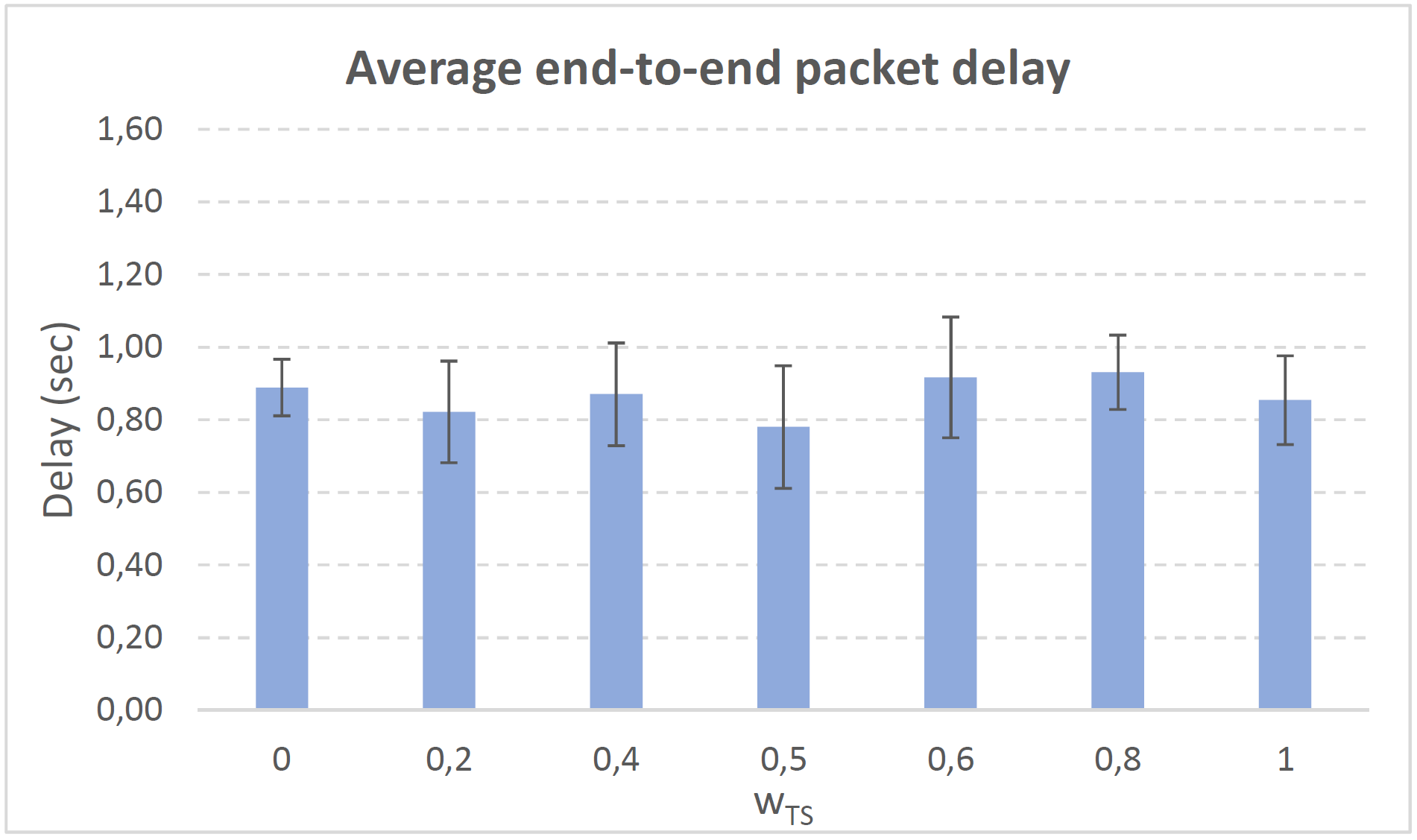}
\caption{Average end-to-end packet delay. Scenario with low social media interaction (TS normal distribution with $\mu=2$ and $\sigma=1$). Nodes' density = 200 nodes/km$^2$.}
\label{fig:200_mu2_delay}
\end{figure}

As we can see, we get the lowest packet losses  (17.5$\%$) when $w_{TS}=0$, i.e. when we select the path with best QoS performance. The highest packet losses (23$\%$) is obtained when $w_{TS}=1$, i.e. when we do not consider QoS performance of the paths but just their TS values. We can consider a good trade-off when $w_{TS}=0.6$ as we give a good importance to the tie strength parameter with not a big increase in the packet losses (18.4$\%$). Regarding the average end-to-end packet delay, for $w_{TS}=0.6$ it is near 0.9 sec and very similar for $w_{TS}=0$, see Figure \ref{fig:200_mu2_delay}.

\subsection{Results for a high social media interaction. TS normal distribution with mean $\mu=3$, standard deviation $\sigma=1$.}

For a scenario with a high social media interaction between users, the tie strength (TS) between consecutive nodes in the forwarding nodes from source to destination is going to be more significant in the forwarding decision and will have a higher impact in the $\overline{TS}$ measured in the simulation.

\subsubsection{High social media interaction. Node's density = 100 nodes/km$^2$}

\begin{figure}
\centering
\includegraphics[height=2in, width=3in]{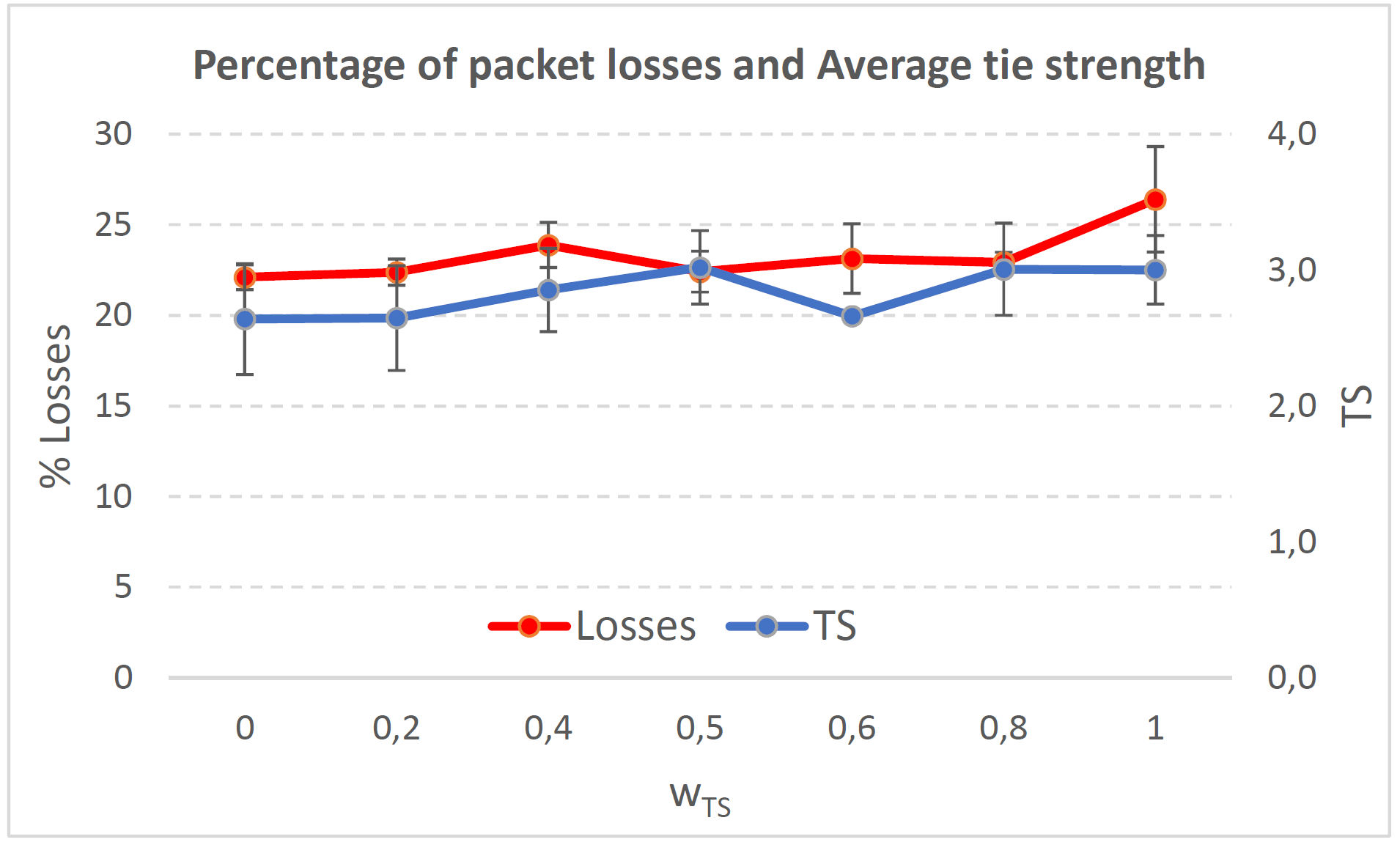}
\caption{Percentage of packet losses and average tie strength of the forwarding path. Scenario with high social media interaction (TS normal distribution with $\mu=3$ and $\sigma=1$). Nodes' density = 100 nodes/km$^2$.}
\label{fig:100_mu3_losses_TS}
\end{figure}

\begin{figure}
\centering
\includegraphics[height=2in, width=3in]{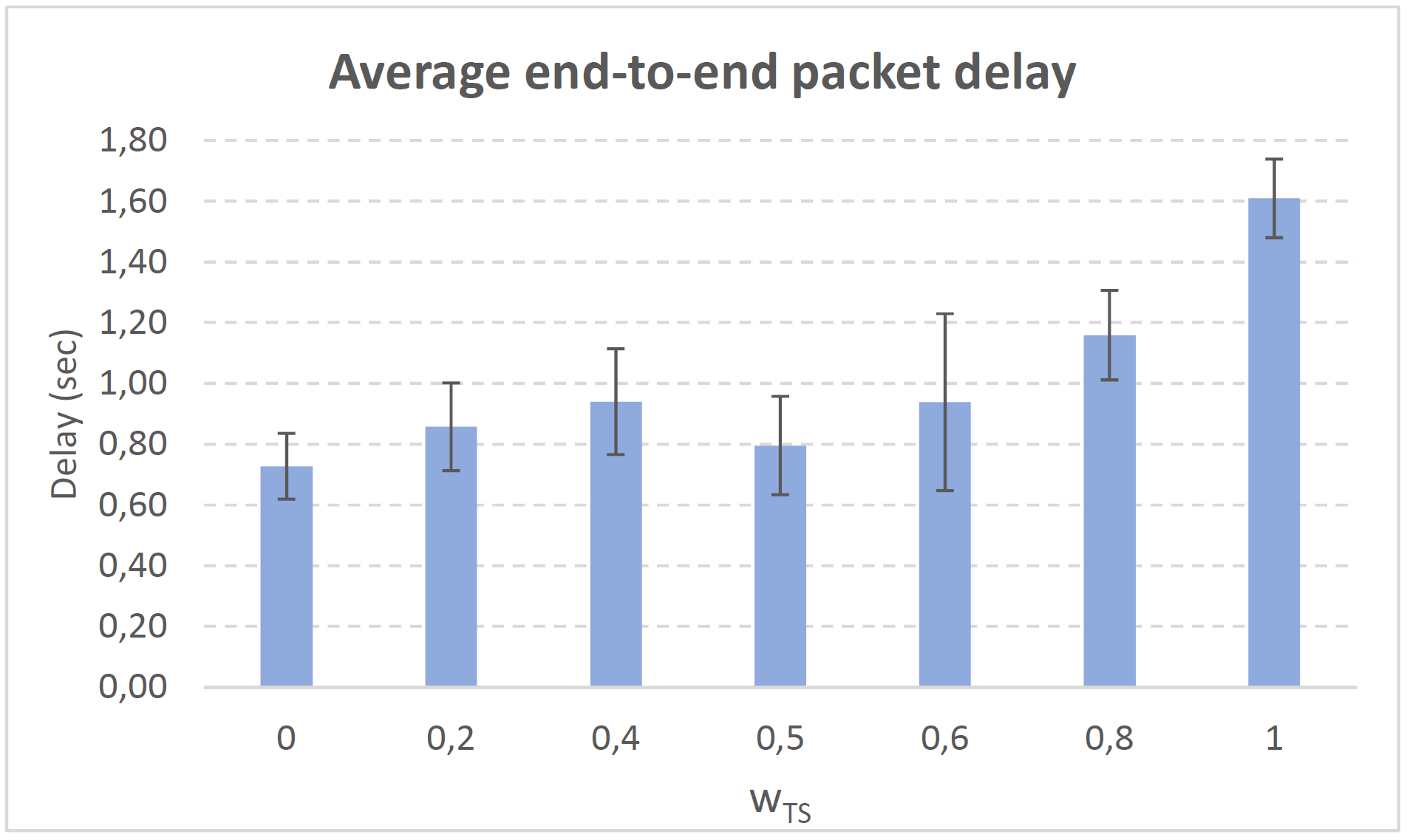}
\caption{Average end-to-end packet delay. Scenario with high social media interaction (TS normal distribution with $\mu=3$ and $\sigma=1$). Nodes' density = 100 nodes/km$^2$.}
\label{fig:100_mu3_delay}
\end{figure}

Figure \ref{fig:100_mu3_losses_TS} shows a percentage of packet losses in the range 22\% (when $w_{TS}=0$, i.e. only QoS parameters are considered to select the forwarding path, see Eq. (\ref{eq:multimetric_score})) to 26\% (when $w_{TS}=1$, i.e. only TS parameter are considered). Between those $w_{TS}=$ values packet losses fluctuates around 23\%. The $\overline{TS}$ measured in the forwarding paths established during simulations varies around 3, see blue line in Fig. \ref{fig:100_mu3_losses_TS}, slightly increasing as $w_{TS}=$ grows since we are giving more importance to the social confidence in front of the QoS offered by the forwarding path. The average end-to-end packet delay ranges between 0.7 and 1.2 sec increasing while $w_{TS}$ grows and reaching 1.6 sec for $w_{TS}=1$. To sum up, a proper value for $w_{TS}=$ could be 0.8, that well balances a trade-off with low packet losses (23\%), low packet delay (0.94 sec) and high $\overline{TS}=2.6$. 

\subsubsection{High social media interaction. Node's density = 200 nodes/km$^2$}

\begin{figure}
\centering
\includegraphics[height=2in, width=3in]{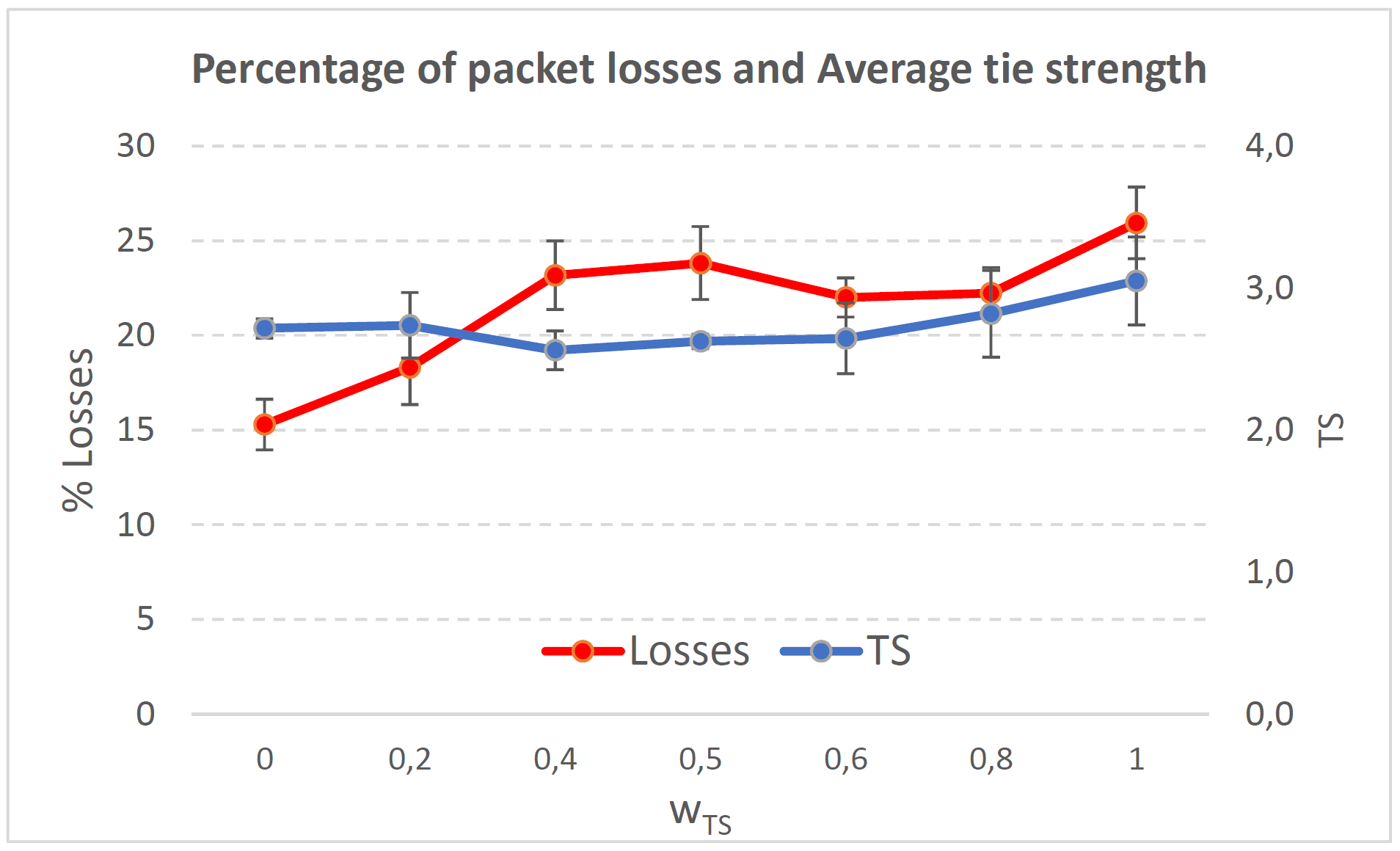}
\caption{Percentage of packet losses and average tie strength of the forwarding path. Scenario with high social media interaction (TS normal distribution with $\mu=3$ and $\sigma=1$). Nodes' density = 200 nodes/km$^2$.}
\label{fig:200_mu3_losses_TS}
\end{figure}

\begin{figure}
\centering
\includegraphics[height=2in, width=3in]{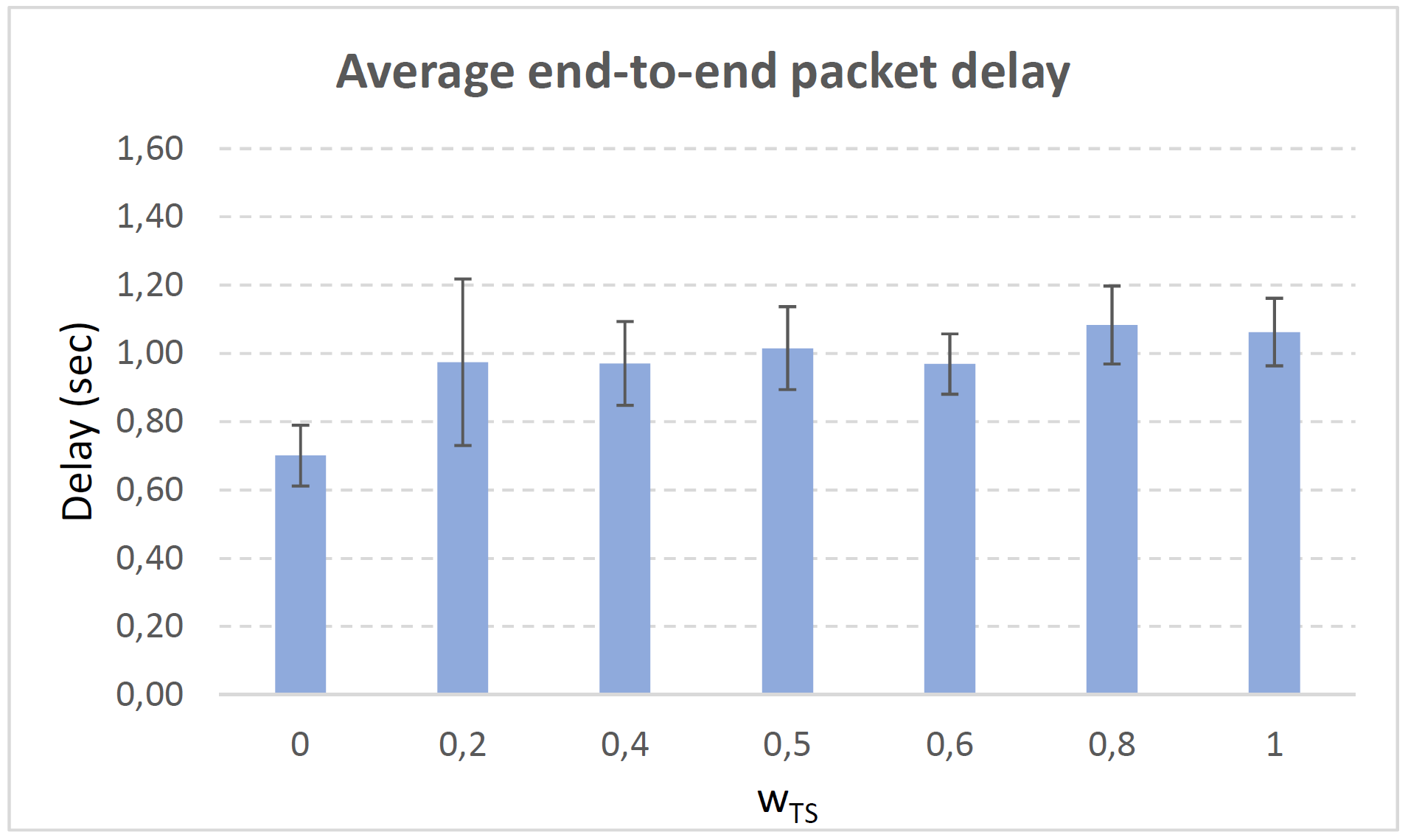}
\caption{Average end-to-end packet delay. Scenario with high social media interaction (TS normal distribution with $\mu=3$ and $\sigma=1$). Nodes' density = 200 nodes/km$^2$.}
\label{fig:200_mu3_delay}
\end{figure}

In a high nodes' density scenario, network connectivity improves and there are more neighboring nodes to be chosen to establish the forwarding path. Furthermore, when the social media interaction between users is high, the  $w_{TS}=$ parameter is more significant in the path selection process, see Eq. (\ref{eq:multimetric_score}). Consequently, we clearly see how the packet losses curve grows as $w_{TS}=$ grows, see red line in Fig. \ref{fig:200_mu3_losses_TS}. Also, the $\overline{TS}$ measured grows, see blue line in the same figure. 

In Fig. \ref{fig:200_mu3_losses_TS}, we can see that the lowest packet loss percentage (15.3$\%$) takes place for $w_{TS}=0$, i.e. when we select the path with best QoS performance. The highest packet loss percentage (26$\%$) occurs for $w_{TS}=1$, i.e. when we do not consider QoS performance of the paths but only their TS values. A good trade-off could be 0.8, with acceptable packet losses (22.2\%), low packet delay (1 sec) and high $\overline{TS}=2.8$. Nonetheless, specially for applications sensitive to packet losses, a better option is just considering the best QoS path ($w_{TS}=0$) which produces 15\% of packet losses. In that case, the measured $\overline{TS}=2.7$ is almost the same. The reason is that in this scenario all the nodes have a high social interaction, so the best strategy to select the forwarding nodes is considering the QoS parameters, since the $\overline{TS}$ will be high for any $w_{TS}$.

\subsection{Results for a very high social media interaction. TS normal distribution with mean $\mu=4$, standard deviation $\sigma=1$.}

For a scenario with a very high social media interaction between users, the tie strength (TS) between consecutive nodes in the forwarding nodes from source to destination will be even more decisive in the forwarding operation and will produce a higher $\overline{TS}$ measured in the simulation.

\subsubsection{Very high social media interaction. Node's density = 100 nodes/km$^2$}

\begin{figure}
\centering
\includegraphics[height=2in, width=3in]{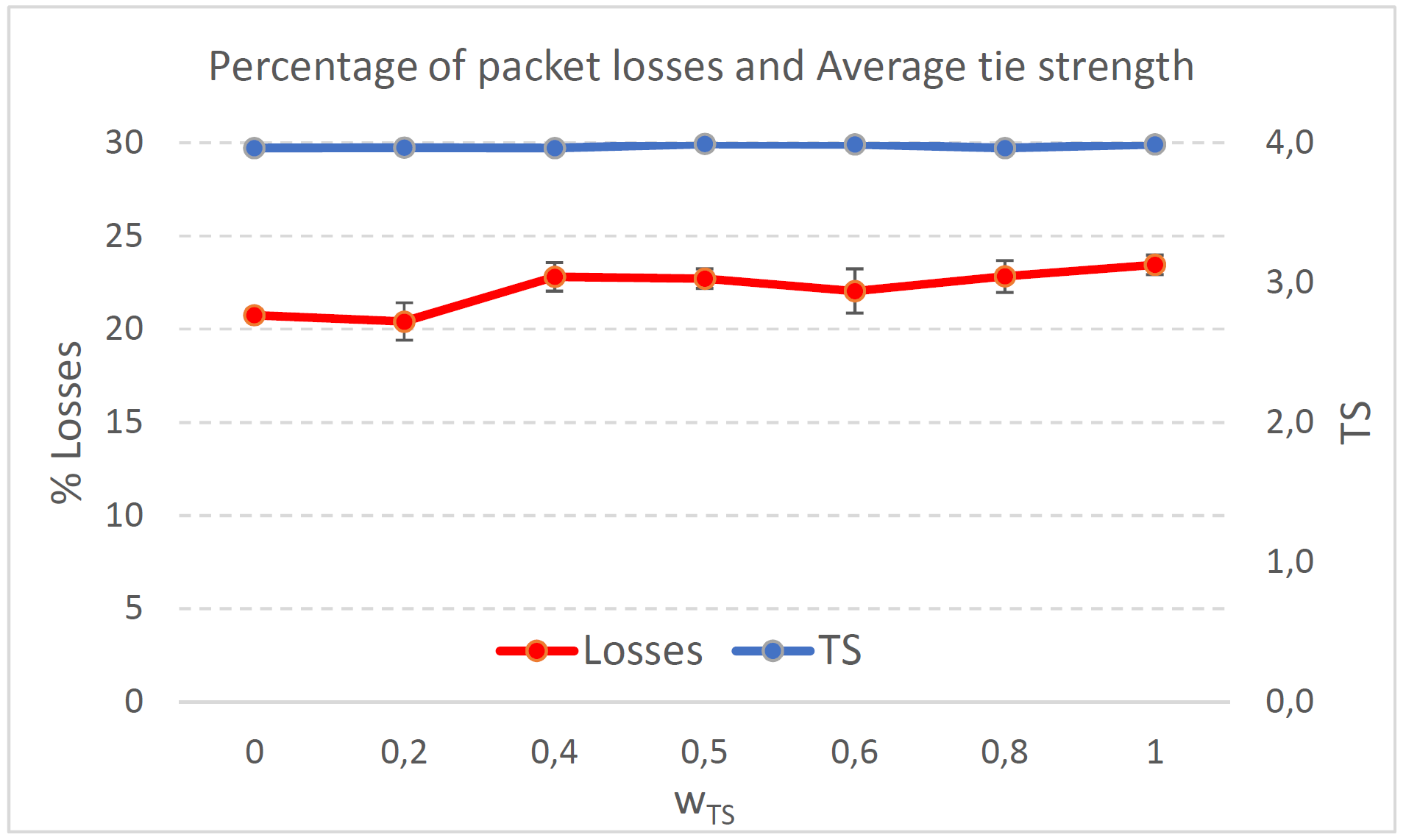}
\caption{Percentage of packet losses and average tie strength of the forwarding path. Scenario with very high social media interaction (TS normal distribution with $\mu=4$ and $\sigma=1$). Nodes' density = 100 nodes/km$^2$.}
\label{fig:100_mu4_losses_TS}
\end{figure}

\begin{figure}
\centering
\includegraphics[height=2in, width=3in]{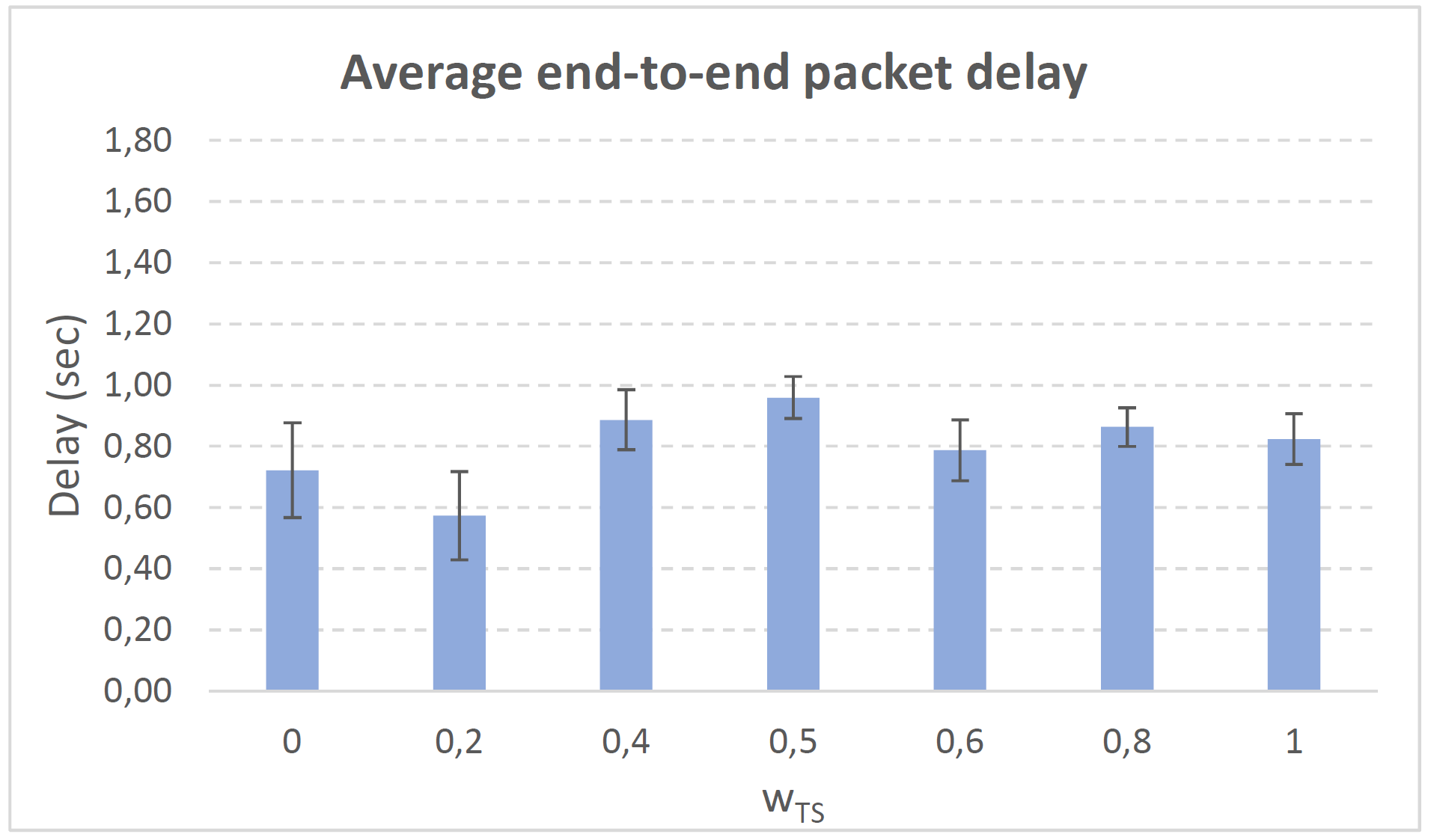}
\caption{Average end-to-end packet delay. Scenario with very high social media interaction (TS normal distribution with $\mu=4$ and $\sigma=1$). Nodes' density = 100 nodes/km$^2$.}
\label{fig:100_mu4_delay}
\end{figure}

\begin{figure}
\centering
\includegraphics[height=2in, width=3in]{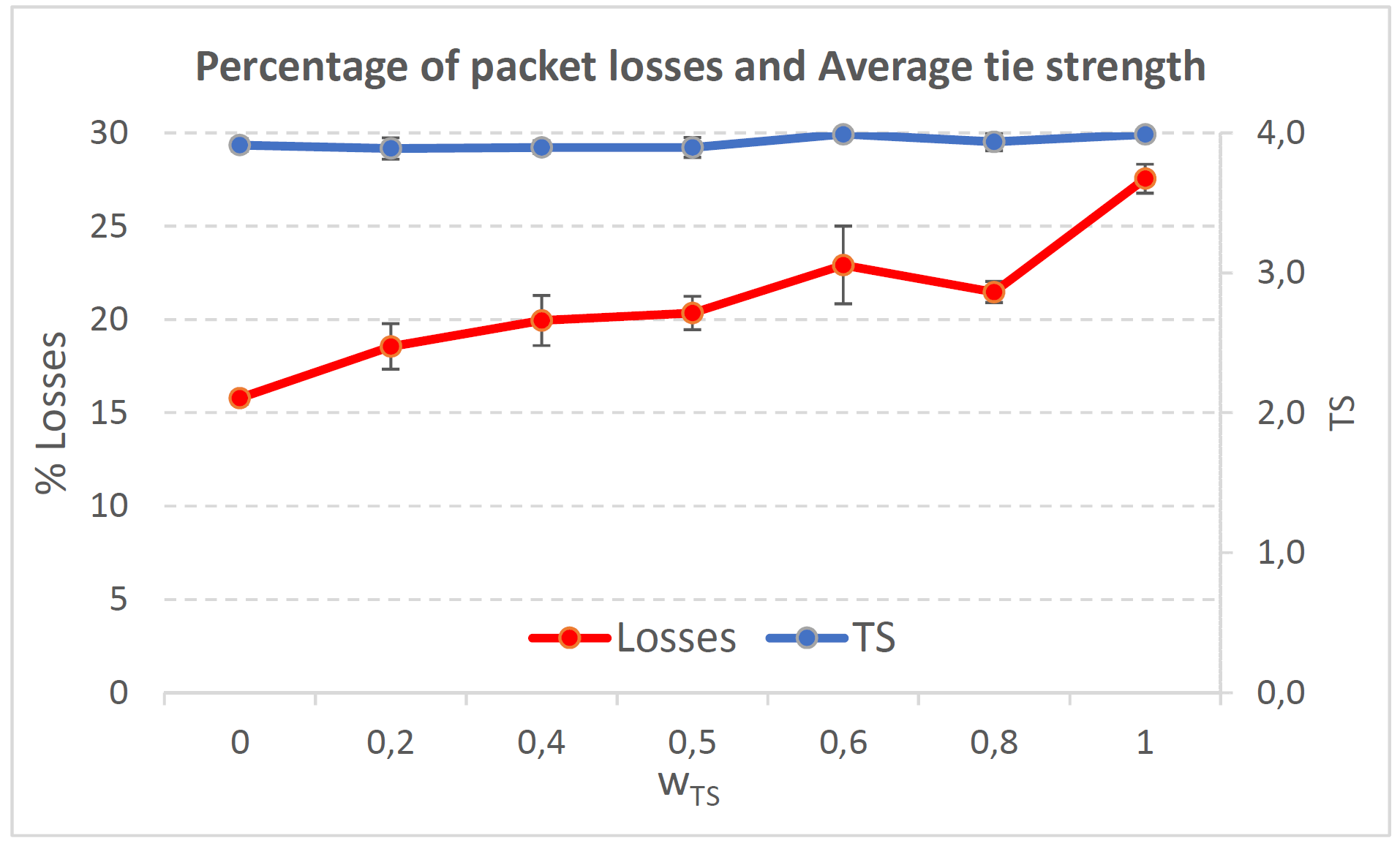}
\caption{Percentage of packet losses and average tie strength of the forwarding path. Scenario with very high social media interaction (TS normal distribution with $\mu=4$ and $\sigma=1$). Nodes' density = 200 nodes/km$^2$.}
\label{fig:200_mu4_losses_TS}
\end{figure}

\begin{figure}
\centering
\includegraphics[height=2in, width=3in]{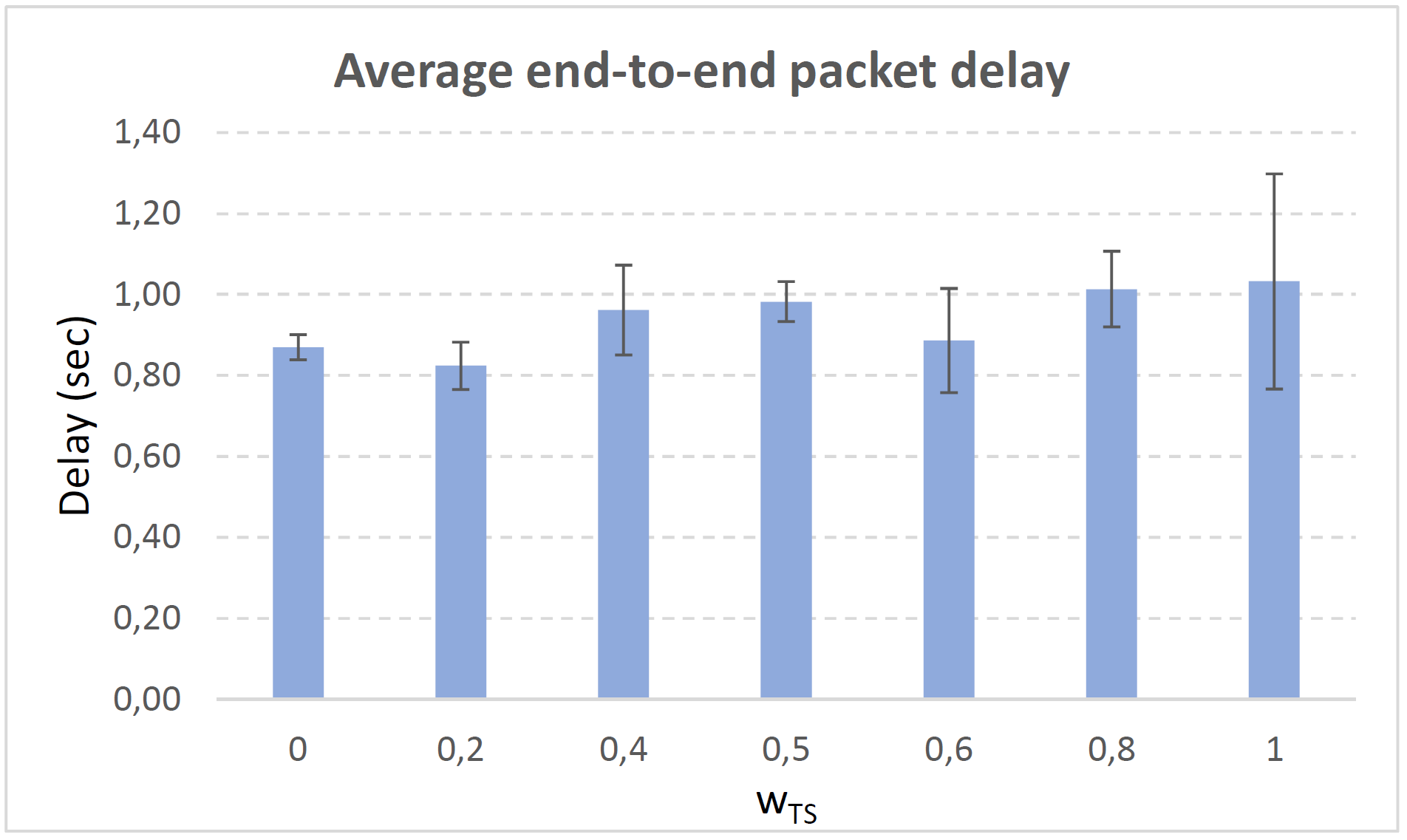}
\caption{Average end-to-end packet delay. Scenario with very high social media interaction (TS normal distribution with $\mu=4$ and $\sigma=1$). Nodes' density = 200 nodes/km$^2$.}
\label{fig:200_mu4_delay}
\end{figure}

Figure \ref{fig:100_mu4_losses_TS} shows the average percentage of packet losses. As we can see, we get the lowest value (around 20$\%$) for $w_{TS}=0$ (i.e.,  we select the path with best QoS performance), and the highest value (around 23$\%$) for $w_{TS}=1$ (i.e., when we just consider the TS values of the paths). Again, there is no much variation in the values when we vary the weight $w_{TS}$, since this density of nodes all the paths have more or less the same QoS performance. Also, for this low
$\sigma=1$ (compared to the average $\mu=4$) nodes show very similar high social media interaction with the other nodes. Therefore, the best $w_{TS}$ is 0, since the $\overline{TS}$ measured is very near to 4 in all the cases.

\subsubsection{Very high social media interaction. Node's density = 200 nodes/km$^2$}

The same conclusion as in the previous case also applies in this case which even can be seen more clearly. For this low $\sigma=1$ (compared to the average $\mu=4$) nodes show very similar TS values regarding the other nodes. Therefore, the best $w_{TS}$ is 0, since the $\overline{TS}$ measured is very near to 4 in all the cases. The lowest percentage of packet losses (15$\%$) is obtained for $w_{TS}=0$ (i.e., when we select the path with best QoS performance), which in this case produces a $\overline{TS}$=4.

Concluding, for this higher nodes' density, it is specially important to consider the QoS performance of the available paths to select the best one. Furthermore, since the TS is very high in all the nodes, the selected path will offer a high $\overline{TS}$ for any  $w_{TS}$.

\section{Conclusions and Future Work}
\label{sec:conclusions}
Providing real-time multimedia services over MANETs represent a very promising and attractive topic due to its interesting applications and challenges. In this work, we have presented a new proposal named \textit{QoS-aware and social-aware multimetric routing protocol for video-streaming services over MANETs} (QSMVM). We have shown that we can take advantage of the growing omnipresence of the social web technologies through the information of interaction among users in social networks to design routing protocols for MANETs. Our goal is to achieve a good trade-off between QoS and confidence in the forwarding path established by our self-configured dynamic framework to offer video-streaming services over MANETs.

Simulation results show the benefits of our approach in terms of QoS provided (percentage of packet losses and average end-to-end packet delay) and average tie strength measured in the forwarding path. In this work, we have modeled a social tie strength (TS) that follows a normal distribution with a given average  $\mu$ and a standard deviation $\sigma$. Benefices are specially noticeable under a high nodes' density (200 $nodes/km^2$) with a high social media interaction between enough pairs of users. We have analysed the effect of the weight of the social confidence provided by the forwarding path ($w_{TS}$ in Eq. (\ref{eq:multimetric_score})) in front of the weight of the QoS provided by the forwarding path ($w_{QoS}$ in Eq. (\ref{eq:multimetric_score})), looking to the suitable $w_{TS}$ value that balances a good trade-off between QoS and social confidence. The goal is to give a high importance to the level of trust among users forming the forwarding paths through which packets are routed, although till a maximum $w_{TS}$ that guarantees that we do not have a big impact in the QoS.

In this article, we have emulated diverse scenarios from (i) scenarios with users without social interaction between users (low tie strength values, e.g. a big city like New York), to (ii) scenarios with high social interaction between users (high tie strength values, e.g. a University Campus). We have modeled the tie strength (TS) values with a normal distribution with means $\mu$ = 1, 2, 3 and 4 and standard deviation of $\sigma$ = 1. A good suggestion for future work would be to carry out simulations in a scenario with real values of the tie strength gathered from social networks such as Twitter or Facebook.

In this work, we have given the same weight $w_{QoS}$ to all the QoS parameters (see Eq. (\ref{eq:multimetric_score})). Thus, in this work we have analysed the trade-off between a global measure of the QoS and the $\overline{TS}$. That is, we have given the same importance to all QoS parameters (available bandwidth, packet losses, packet delay, delay jitter, hop-count distance to destination, reliability metric, mobility metric). We are currently working on the design of different weights to those QoS metrics, e.g. giving higher values to the percentage of packet losses or packet delay so that they have a higher impact on the selection of forwarding paths, which may be interesting depending on the applications.


\begin{thebibliography}{}

\bibitem[bon, ]{bonmotion}
Bonnmotion: Tool to generate mobility scenarios for mobile ad hoc networks.
\newblock {\em http://sys.cs.uos.de/bonnmotion/}.

\bibitem[IEE, ]{IEEE}
Ieee 802.11e standard with quality of service enhancements.
\newblock {\em http://standards.ieee.org/getieee802/download/802.11e-2005.pdf}.

\bibitem[ns2, ]{ns2}
The network simulator, ns-2.
\newblock {\em http://nsnam.isi.edu/nsnam/}.

\bibitem[{Aguilar Igartua} et~al., 2020]{INRISCO}
{Aguilar Igartua}, M., {Mendoza}, F.~A., {Díaz Redondo}, R.~P., {Vicente}, M. I.~M., {Forné}, J., {Campo}, C., {Fernández-Vilas}, A., {De La Cruz Llopis}, L.~J., {García-Rubio}, C., {López}, A.~M., {Mezher}, A.~M., {Díaz-Sánchez}, D., {Cerezo-Costas}, H., {Rebollo-Monedero}, D., {Arias-Cabarcos}, P., and {Rico-Novella}, F.~J. (2020).
\newblock Inrisco: Incident monitoring in smart communities.
\newblock {\em IEEE Access}, 8:72435--72460.

\bibitem[{Bader} and {Alouini}, 2017]{QoS_disaster}
{Bader}, A. and {Alouini}, M. (2017).
\newblock Mobile ad hoc networks in bandwidth-demanding mission-critical applications: Practical implementation insights.
\newblock {\em IEEE Access}, 5:891--910.

\bibitem[{Bhardwaj} and {El-Ocla}, 2020]{QoS_Bhard}
{Bhardwaj}, A. and {El-Ocla}, H. (2020).
\newblock Multipath routing protocol using genetic algorithm in mobile ad hoc networks.
\newblock {\em IEEE Access}, 8:177534--177548.

\bibitem[{Boukerche}, 2009]{Azzedine_book}
{Boukerche}, A. (2009).
\newblock {\em Routing Protocols in Intermittently Connected Mobile Ad Hoc Networks and Delay‐Tolerant Networks}, pages 219--250.

\bibitem[{Chen} et~al., 2013]{integrated_social_QoS}
{Chen}, I., {Jia Guo}, {Bao}, F., and {Cho}, J. (2013).
\newblock Integrated social and quality of service trust management of mobile groups in ad hoc networks.
\newblock In {\em 9th International Conference on Information, Communications Signal Processing}, pages 1--5.

\bibitem[Choi et~al., 2017]{choi2017wom}
Choi, Y.~K., Seo, Y., and Yoon, S. (2017).
\newblock E-wom messaging on social media.
\newblock {\em Internet Research}.

\bibitem[{Dressler}, 2008]{Falko_book}
{Dressler}, F. (2008).
\newblock {\em Mobile Ad Hoc and Sensor Networks}, pages 67--94.

\bibitem[Granovetter, 1973]{Granovetter73}
Granovetter, M.~S. (1973).
\newblock The strength of weak ties.
\newblock {\em The American Journal of Sociology}, 78(6):1360--1380.

\bibitem[{Greco} et~al., 2012a]{QoS_Greco}
{Greco}, C., {Cagnazzo}, M., and {Pesquet-Popescu}, B. (2012a).
\newblock Low-latency video streaming with congestion control in mobile ad-hoc networks.
\newblock {\em IEEE Transactions on Multimedia}, 14(4):1337--1350.

\bibitem[{Greco} et~al., 2012b]{QoS_video3}
{Greco}, C., {Cagnazzo}, M., and {Pesquet-Popescu}, B. (2012b).
\newblock Low-latency video streaming with congestion control in mobile ad-hoc networks.
\newblock {\em IEEE Transactions on Multimedia}, 14(4):1337--1350.

\bibitem[Huang et~al., 2018]{huang2018will}
Huang, H., Dong, Y., Tang, J., Yang, H., Chawla, N.~V., and Fu, X. (2018).
\newblock Will triadic closure strengthen ties in social networks?
\newblock {\em ACM Transactions on Knowledge Discovery from Data (TKDD)}, 12(3):1--25.

\bibitem[Igartua et~al., 0112]{QoS_video2}
Igartua, M.~A., de~la Cruz~Llopis, L.~J., Frías, V.~C., and Gargallo, E.~S. (20112).
\newblock A game-theoretic multipath routing for video-streaming services over mobile ad hoc networks.
\newblock {\em Computer Networks}, 55(13):2985--3000.

\bibitem[Igartua et~al., 2012]{QoS_video1}
Igartua, M.~A., Frías, V.~C., de~la Cruz~Llopis, L.~J., and Gargallo, E.~S. (2012).
\newblock Dynamic framework with adaptive contention window and multipath routing for video-streaming services over mobile ad hoc networks.
\newblock {\em Telecommunications Systems}, 49(4):379–390.

\bibitem[{Jeong} and {Shin}, 2018]{HC_social_MANET}
{Jeong}, C. and {Shin}, W. (2018).
\newblock Network-decomposed hierarchical cooperation in ad hoc networks with social relationships.
\newblock {\em IEEE Transactions on Wireless Communications}, 17(11):7606--7619.

\bibitem[Kone and Nandi, 2006]{AQR_QoS_Video_MANET}
Kone, V. and Nandi, S. (2006).
\newblock Qos constrained adaptive routing protocol for mobile adhoc networks.
\newblock In {\em 9th International Conference on Information Technology, ICIT}, pages 1--6.

\bibitem[Mezher et~al., 2015a]{GT_video_MANET_Ahmad}
Mezher, A., Igartua, M., De~la Cruz~Llopis, L.J.and~Segarra, E., Tripp-Barba, C., Urquiza-Aguiar, L., Forné, J., and Gargallo, E. (2015a).
\newblock A multi-user game-theoretical multipath routing protocol to send video-warning messages over mobile ad hoc networks.
\newblock {\em Sensors}, 15:9039--9077.

\bibitem[Mezher et~al., 2015b]{Ahmad_Jitel_2015}
Mezher, A.~M., Paredes, C.~I., Tripp-Barba, C., and Igartua, M.~A. (2015b).
\newblock A dynamic multimetric weights distribution in a multipath routing protocol using video-streaming services over manets.
\newblock {\em XII Jornadas de Ingeniería Telemática, JITEL}.

\bibitem[{Paramasivan} et~al., 2015]{QoS_Paramasivan}
{Paramasivan}, B., {Prakash}, M. J.~V., and {Kaliappan}, M. (2015).
\newblock Development of a secure routing protocol using game theory model in mobile ad hoc networks.
\newblock {\em Journal of Communications and Networks}, 17:75--83.

\bibitem[{RFC 4728}, 2007]{RFC4728}
{RFC 4728} (2007).
\newblock The dynamic source routing protocol (dsr) for mobile ad hoc networks for ipv4,.

\bibitem[Servia-Rodr{\'\i}guez et~al., 2015]{servia2015evolution}
Servia-Rodr{\'\i}guez, S., Noulas, A., Mascolo, C., Fern{\'a}ndez-Vilas, A., and D{\'\i}az-Redondo, R.~P. (2015).
\newblock The evolution of your success lies at the centre of your co-authorship network.
\newblock {\em PloS one}, 10(3):e0114302.

\bibitem[Servia-Rodríguez et~al., 2014]{Servia13}
Servia-Rodríguez, S., Díaz-Redondo, R.~P., Fernández-Vilas, A., Blanco-Fernández, Y., and Pazos-Arias, J.~J. (2014).
\newblock A tie strength based model to socially-enhance applications and its enabling implementation: mysocialsphere.
\newblock {\em Expert Systems with Applications}, 41(5):2582 -- 2594.

\bibitem[Song et~al., 2017]{song2017whose}
Song, T., Yi, C., and Huang, J. (2017).
\newblock Whose recommendations do you follow? an investigation of tie strength, shopping stage, and deal scarcity.
\newblock {\em Information \& Management}, 54(8):1072--1083.

\bibitem[Streijl et~al., 2016]{MOS}
Streijl, R.~C., Winkler, S., and Hands, B. D.~S. (2016).
\newblock Mean opinion score (mos) revisited: methods and applications, limitations and alternatives.
\newblock {\em Multimedia Systems}, 22(2):213--227.

\bibitem[{Taha} et~al., 2017]{QoS_Taha}
{Taha}, A., {Alsaqour}, R., {Uddin}, M., {Abdelhaq}, M., and {Saba}, T. (2017).
\newblock Energy efficient multipath routing protocol for mobile ad-hoc network using the fitness function.
\newblock {\em IEEE Access}, 5:10369--10381.

\bibitem[{Takahata} et~al., ]{data_rate_video_manet}
{Takahata}, K., {Uchida}, N., and {Shibata}, Y.
\newblock Optimal data rate control for video stream transmission over wireless network.
\newblock In {\em 18th International Conference on Advanced Information Networking and Applications, AINA 2004}, volume~1, pages 340--345.

\bibitem[Tsai, 2009]{p2p_video_manet}
Tsai, T.-C. (2009).
\newblock Quality-aware multiple backbone construction on multi-interface wireless mesh networks for p2p streaming.
\newblock In {\em 2009 IEEE 6th International Conference on Mobile Adhoc and Sensor Systems}.

\bibitem[{Venkatesh} et~al., 2019]{ANFIS_QoS_Video_MANET}
{Venkatesh}, K., {Nithiyanandam}, N., and {Sivaneshkumara} (2019).
\newblock Anfis based qos-aware routing protocol for video streaming in manets.
\newblock In {\em 2019 IEEE International Conference on Intelligent Techniques in Control, Optimization and Signal Processing, INCOS}, pages 1--6.

\bibitem[{Weng} et~al., 2010]{QoS_Weng}
{Weng}, C., {Chen}, C., {Ku}, C., and {Hwang}, S. (2010).
\newblock A bandwidth-based power-aware routing protocol with low route discovery overhead in mobile ad hoc networks.
\newblock {\em The Computer Journal}, 53(7):969--990.

\end{thebibliography}

\end{document}